\documentclass[twocolumn,amssymb,amsmath,floats,showpacs,superscriptaddress,pre]{revtex4-1}
\usepackage{amsmath,amssymb,bm}
\usepackage{graphicx}

\usepackage{comment}
\usepackage[hidelinks]{hyperref}

\begin{document}

\title{Solution of the explosive percolation quest:
\\
Scaling functions and critical exponents}
%%\title{Pearson coefficient of strongly correlated growing trees approaches zero}

\author{R. A. da Costa}
\affiliation{Departamento de F{\'\i}sica, I3N, Universidade de Aveiro, 3810-193 Aveiro,
Portugal}
\author{S. N. Dorogovtsev}\email{sdorogov@ua.pt}
\affiliation{Departamento de F{\'\i}sica, I3N, Universidade de Aveiro, 3810-193 Aveiro,
Portugal}
\affiliation{A. F. Ioffe Physico-Technical Institute, 194021 St. Petersburg, Russia}
%%\affiliation{e-mail:sdorogov@ua.pt}
%%\email{sdorogov@ua.pt}
\author{A. V. Goltsev}
\affiliation{Departamento de F{\'\i}sica, I3N, Universidade de Aveiro, 3810-193 Aveiro,
Portugal}
\affiliation{A. F. Ioffe Physico-Technical Institute, 194021 St. Petersburg, Russia}
\author{J. F. F. Mendes}
\affiliation{Departamento de F{\'\i}sica, I3N, Universidade de Aveiro, 3810-193 Aveiro,
Portugal}

\begin{abstract}

%%The notion of percolation 
%%assumes 
%%means 
Percolation 
%%implies 
refers to the emergence of a giant connected
%%%%(percolating) 
cluster in 
a 
%%large 
disordered 
system 
%%systems 
when the number of 
connections between nodes 
%%links 
exceeds 
%%some 
a critical value. 
%%This transition can 
%%%%also 
%%be treated as a critical point in the process of subsequent 
%%%%interlinking 
%%%%connecting of 
%%%%adding 
%%creating 
%%%%links 
%%connections between pairs of randomly selected nodes. 
The percolation phase transitions
were 
%%commonly 
believed to be continuous until recently when in a new so-called
``explosive percolation'' problem for 
%%a Metropolis-like algorithm 
%%a local 
a competition 
%%an optimization 
driven process,  
%%evolution,
a discontinuous phase transition was 
%%first 
reported. 
%%%%based on 
%%a 
%%%%computer experiments  
%%%%\cite{Achlioptas:ads09,Cho:ckp09,Radicchi:rf09,Ziff:z09}. 
%%[1--4]. 
%%Nonetheless, 
%%However, 
%%Although 
%%the 
The 
%%direct 
analysis of 
%%numerical solution of the 
%%first million 
evolution equations for this process showed however that this 
%%strange 
transition is
actually continuous though 
%%demonstrating 
with 
%%unique 
%%surprising 
%%%%quite unusual 
%%features and 
%%%%values of 
surprisingly tiny critical exponents. 
%%%%\cite{daCosta:ddgm10}. 
%%Afterwards, this continuity was proven mathematically \cite{xxx}.   
%%though with unique properties 
%%[5]. 
%%, current opinion still favors discontinuity. 
%%this result convinced yet  
%%only a 
%%small minority of 
%%few researchers. 
%%Despite these results, experimental arguments in favor of the discontinuity of transitions of this kind continue to rage making explosive percolation one of the 
%%hottest 
%%highly 
%%discussed issues in modern statistical physics. 
%%%%One should stress that the 
%%main 
%%%%most exciting problem 
%%The 
%%%%most exciting, 
%%main problem, however, 
%%%%is not the continuity of the ``explosive percolation'' transition but rather its 
%%unusual 
%%%%unique features, sharply distinct from standard percolation.  
%%Despite these results, ------ stayed uncovered... 
%%one  which results in the 
%%%%wide 
%%hot discussion of this issue. 
%%and emerge are still emerged and discussed.    
%%%%The 
%%real 
%%%%question is: what is the 
%%real 
%%%%nature of this strange 
%%misterious 
%%%%transition?  
%%%%Here we present a complete solution of this intriguing quest. 
%%Here we explain completely the nature of this strange transition. 
%%Namely, for 
For a wide class of representative models, we develop a strict scaling theory of this 
exotic 
%strange 
transition 
which 
%%explains the continuous nature of this transition and 
%%gives 
provides 
%%enables us to obtain 
the full set of scaling functions and critical exponents. 
%%for each of the models with any desired precision. 
%%, proving that the transition is second-order. 
This theory indicates the relevant order parameter and susceptibility for the problem, and 
explains the continuous nature of this 
%%second-order 
transition and its unusual 
%%features 
properties.  
%%for these processes.    
%%This theory explains the continuous nature of this 
%%%%second-order 
%%transition and its unusual 
%%%%features 
%%properties and indicates the relevant order parameter and susceptibility. 
%%%%for these processes.    

\end{abstract}


\maketitle

%%\email{sdorogov@fis.ua.pt}

%%%%%%%\email{goltsev@fis.ua.pt}

%%%%%%%\email{jfmendes@fis.ua.pt}

%%\date{}

%%\cite{Dorogovtsev:dgm08,Goltsev:gdm08} \cite{Newman:n02b} 
%%\cite{Boguna:bpv03} \cite{Barrat:bp05} 
%%\cite{Krapivsky:kr01} 
%%\cite{Dorogovtsev:dm01} 
%%\cite {Dorogovtsev:dms00} \cite{Krapivsky:krl00}  
%%\cite{Newman:n03} 
%%\cite{Pastor-Satorras:pvv01} \cite{Maslov:ms02} 

%%The 
The percolation phase transition is one of the central issues 
%%of the physics of 
for disordered systems 
%%in condensed matter and statistical mechanics 
%%, including diverse disordered lattices and random networks 
\cite{Stauffer:sa-book94,Stauffer:s79,Dorogovtsev:dgm08,Dorogovtsev:d-book10}.   
%%points for understanding is one of fundamental points in disordered systems the physics  and networks. 
%%chance of the structure \cite{xxx,xxx,xxx}. 
%%In the simplest example, 
%%%%The classical random graphs provide 
%%perhaps 
%%one of the 
%%the simplest 
%%%%a basic example of this transition. Start with an infinite 
%%large (actually infinite) 
%%%%set of $N\to\infty$ unconnected nodes, and at each step add a link between two uniformly randomly
%%%%chosen nodes. When the relative number of links $t=L/N$ in this graph exceeds
%%%%the threshold $t_c=1/2$, the graph 
%%includes 
%%%%has a 
%%%%giant cluster (in other words,
%%%%giant connected component 
%%%%or 
%%%%(percolation cluster) containing a finite fraction
%%%%$S=N_{\text{GC}}
%%{\text{giant}}
%%%%/N$ of all nodes. 
%%One can show that this process can be reversed and that the percolation transition in this system can be treated as equilibrium. 
%%%%This process can be reversed and so the percolation transition in this system 
%%can be treated as 
%%%%is actually equilibrium. 
%%In the neighbourhood of the critical point, $S \propto (t-t_c)$.  
%%Importantly, the 
Phase transitions in 
%%%%this and other 
classical percolation problems are very well known to be continuous, 
that is, 
%%i.e. 
the relative size of the percolation cluster $S$, which is the order parameter for these models,  
%%for these problems,  
%%the percolation cluster relative size 
%%$S$, which is the order parameter for these problems, 
emerges continuously,
without a jump at the percolation threshold. 
%%%%critical point. 
%%%%In particular, for the classical random graph model described above, the relative size of the giant connected component $S \propto (t-t_c)$ in the neighbourhood of the percolation threshold, that is the critical exponent $\beta$, $S \propto (t-t_c)^\beta$, is $1$. 
%%The fact that the phase transitions in percolation problems (apart of special problems such as bootstrap percolation and $k$-cores \cite{xxx}) are continuous was generally accepted until recently. 
%%It was generally accepted that the phase transitions in percolation problems are   
%%%%Furthermore, 
%%Importantly, 
%%One should note that the 
%%continuity of 
%%%%the fact that 
%%classical 
%%The continuity of ordinary percolation phase transitions 
%%are continuous 
%%directly leads to a power-law distribution of cluster sizes at the percolation threshold and to a set of standard scaling properties and relations. 
As a continuous phase transition, the ordinary percolation transition is characterized by a power-law distribution of cluster sizes at the percolation threshold and a set of standard scaling properties and relations. 
This common understanding of percolation 
%%however 
was shaken by 
%%the 
work \cite{Achlioptas:ads09} that reported 
%%reporting 
%%in which 
a discontinuous 
%%The continuity of the classical percolation transitions is ... 
%%A 
%%%%principally 
%%sharply different 
%%very distinct, at first sight, 
percolation phase transition 
%%was reported 
in models whose 
%%systems with 
%%another 
%%a new type of evolution 
evolution was driven by local optimization algorithms. 
Based on a computer experiment for a $512\,,000$ node system \cite{Achlioptas:ads09}, it was concluded 
%%in Ref.~\cite{Achlioptas:ads09} 
that the percolation transition for these 
%%apparently 
irreversible processes 
%%(one cannot invent a reverse process to this one) 
is discontinuous, and that is why this kind of percolation was termed ``explosive percolation''. 
%%%%The conclusion was that the delay of the transition 
%%due to the prevalent merging of small clusters in this scheme 
%%%%results in its discontinuity. 
%%%%a discontinuous, 
%%%%``violent'' 
%%transition. 
This conclusion was 
%%immediately 
supported by a number of simulations of 
%%this and similar 
models of this kind \cite{Cho:ckp09,Radicchi:rf09,Ziff:z09,Cho:ckk10,D'Souza:dm10,Radicchi:rf10,Ziff:z10,Manna:mc11,Araujo:aaz11,Friedman:fl09}. 
%%(Note that in 
%%%%In the present paper we discuss only models of this type.    
%%, although some of 
Surprisingly, these and other studies, 
%%also 
in addition, reported power-law cluster size distributions at the critical point and scaling features below and above $t_c$ (see Ref.~\cite{Ziff:z10,Radicchi:rf10,Araujo:aaz11,Grassberger:gcb11,Cho:ckn10,Lee:lkp11}), unexpected for discontinuous transitions. 

%%In our paper of 2010 \cite{xxx}, we have 
We 
%%have 
%%already 
%%removed 
resolved this contradiction 
by 
showing 
%%This contradiction was removed after the authors of the present work showed 
that the explosive percolation transition is actually continuous though with a uniquely small critical exponent $\beta$ of the percolation cluster size 
%%, $S \propto (t-t_c)^\beta$ 
\cite{daCosta:ddgm10}. We 
obtained this result by analyzing evolution equations for this process in the infinite system size limit.  
%%Note that in the present paper we discuss only  
%%\cite{xxx}. 
Thanks to the smallness of the exponent $\beta$, the continuous transition looks so ``sharp'' that it is virtually impossible to distinguish it from a discontinuous one 
%%The smallness of the exponent $\beta$ explains why it was  
%%%%a direct observation of the continuous transition in computer experiments virtually 
%%%%that it was 
%%%%quite difficult 
%%virtually impossible to observe the continuous transition directly 
in computer experiments even for very large systems \cite{daCosta:ddgm10}. More recently, 
%%the continuity of this transition 
the fact that this transition is continuous was also 
%%proven mathematically 
supported by mathematicians \cite{Riordan:rw10}. Nonetheless, 
%%speaking in terms 
%%in the sense of physics, 
in the physics sense, 
the quest of the explosive percolation transition 
%%still 
actually has not been yet resolved. 
The 
%%point is that the 
main problem 
%%here 
is 
%%actually not the continuity or discontinuity of the transition 
%%(the original confusing conclusion was simply based on misinterpretation of the results of computer experiment for a very small system; 
%%(in principle, the continuity 
%%%%directly 
%%clearly 
%%%%readily and undoubtedly 
%%%%immediately 
%%followed 
%%is obvious 
%%readily follows from the 
%%observations, namely, the 
%%observed power law at the critical point and scaling above and below the transition),  
%%the reported discontinuity was based) 
%%but 
%%the nature of 
%%rather 
how to explain 
the nature of
 this surprising physical phenomenon. 
%%its 
%%interesting 
%%surprising 
%%unique 
%%features which 
%%just 
%%led to the original confusion. 
%%Indeed, what does mean complete description and understanding of a continuous phase transition from the point of view of a theoretical physicist?  
%%Indeed, for a theoretical physicist, the complete solution of the problem of a given continuous transition (in other words, complete description and understanding of the transition) 
%%%%(in other words, description and understanding ) 
%%means (i) indicating the order parameter and the generalized susceptibility for this transition, (ii) finding the full set of scaling relations, (iii) obtaining the scaling functions and critical exponents in these relations, (iv) finding the upper critical dimension for this transition. 
%%Indeed, for 

Here, for this explosive percolation transition in a wide set of representative models, we fulfill the following program. 
%%which comprises the following issues: 
%%(i) 
%For this transition, 
We indicate 
%%indicating 
the order parameter and the generalized susceptibility,  
%%for this transition, 
%%(ii) 
find the full set of scaling relations and relations between critical exponents,  
%%(iii) 
obtain the scaling functions and critical exponents,  
%%entering these relations 
and get
%%(iv) 
the upper critical dimension 
%%$d_c$ 
%%for this transition 
(that is, the dimension, above which 
%%, $d>d_c$, 
%%recall that for systems of dimensionality $d$ higher than $d_c$, 
a mean-field 
%%theories are 
description valid). 
%%; e.g., for ordinary percolation, $d_c=6$ \cite{xxx}). : 
%%For a physicist, resolving these issues means the complete description and understanding of a given continuous continuous phase transition. 
In short, we develop a scaling theory of this transition.

The main body of this paper is organized as follows. In Sec.~\ref{s1} we give the definition of the considered set of models. In Sec.~\ref{s2} we derive the evolution equations corresponding to those models. Section~\ref{s3} shows the set of scaling relations between critical exponents for this explosive percolation transition. In Sec.~\ref{s4} we indicate the proper order parameter and susceptibility for explosive percolation. Section~\ref{s5} shows the set of hyperscaling relations between critical exponents and spatial dimensions. In Sec.~\ref{s6} we outline the derivation of the equations for the scaling functions and describe their solutions, including the precise values of the critical exponents. In Sec.~\ref{s7} we discuss and summarize the results of this paper. In Appendices we give the details of our theory.

\begin{figure}[t]
%%%%[tbhd]
\begin{center}
\scalebox{0.135}{\includegraphics[angle=0]{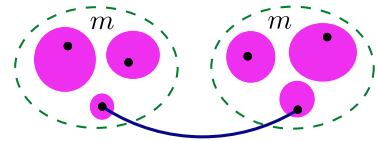}}
\end{center}
\caption{ 
{\bf Illustration of rules in the model of explosive percolation.} 
At each step, two sets of $m$ nodes are chosen at random. Within each set, the node in the smallest cluster is selected, and these two nodes are interconnected. 
%%Here the number $m$ of nodes selected in each of two random samples  is 3.  
} 
\label{f1}
\end{figure}
%%%%%%%%%%%%%%%%%%%%%%%%%%%%%%%%%%%%%%%%%%%%%%%
%%%%%%%%%%%%%%%%%%%%%%%%%%%%%%%%%%%%%%%%%%%%%%%

%%%%%%%%%%%%%%%%%%%%%%%%%%%%%%%%%%%%%%%%
%%%%%%%%%%%%%%%%%%%%%%%%%%%%%%%%%%%%%%%%
%%%%%%%%%%%%%%%%%%%%%%%%%%%%%%%%%%%%%%%%

\section{The models}
\label{s1}

In this work we consider a set of 
%%infinite-dimensional 
models 
of evolving networks which generalizes ordinary percolation on classical random graphs. The number $N$ of nodes is fixed. At each time step a new link connecting two nodes is added to the network. The evolution rules define how these nodes 
%%are 
must be selected. 
%%%%The set of infinite-dimensional models of evolving networks which we consider in this work is as follows. The number of nodes $N$ in the networks does not change during the evolution. At each time step a new link connecting two nodes is added to the network. The evolution rules define how these nodes are selected.  
%%In this work we consider the following set of models which naturally generalizes the standard percolation 
%%(see above) 
%%, which was described above, 
%%and leads to treatable evolution equations. 
Initially the network consists of a given set of finite clusters. 
%%, in total, of $N$ nodes. 
For example, these may be $N$ unconnected nodes. 
%%Then we can 
%%%%It is convenient to introduce an effective time $t\equiv L/N$, where $L$ is the number of links in the network. 
%%The evolution starts from a given set of clusters which consist of 
At each 
%%time 
step 
%%we 
sample two times (see Fig.~\ref{f1}):  

\noindent
(i) choose $m\geq 1$ nodes uniformly at random and compare the clusters to which these nodes belong; 
%%of these $m$ nodes, 
select the 
%%one 
node within  
%%select that node of the $m$ ones, which belongs 
the smallest of these clusters; 
\\
(ii) similarly choose the second 
%%sample 
sampling of $m$ nodes and, again, as in (i), select the node belonging to the smallest of the $m$ clusters; 
\\
(iii) add a link between the two selected nodes thus merging the two smallest clusters. 
%%(see Fig.~\ref{f1}). 
%%
\\
%%\noindent 
In particular, if $m=1$, we arrive at 
%%standard 
ordinary percolation, in which at each step two randomly selected nodes are interconnected. 
%%%%On the other hand, if the number $m$ approaches infinity, the pair of the smallest clusters in the network merge at each step, and so the giant connected component will emerge only after all finite clusters will merge with each other and disappear. 
%%%%Let us introduce an effective time $t\equiv L/N$, where $L$ is the number of links in the network. 
%%%%This will happen at $t_c=1$, and in this 
%%and only this 
%%%%%case 
%%, clearly, 
%%%%the relative size of the giant component is indeed discontinuous at the transition, $S(t_c-0)=0$ and $S(t_c+0)=1$. We will show that this is the only situation with discontinuity. 
%%Clearly, this limiting case cannot be called ``explosive'', since the
Importantly, our rules contain the basic element of other explosive percolation models \cite{Achlioptas:ads09,D'Souza:dm10,Nagler:nlt11,Friedman:fl09,Araujo:aaz11} implementing local optimization rules, namely, selection the minimal clusters from a few possibilities. 
%%Moreover, one can show that, when $m\geq 2$, t
For $m>1$, this selection is performed more efficiently than in the original explosive percolation  model, 
%%see Ref.~\cite{daCosta:ddgm10}, 
%%which means that, in our models, 
since, in average, our rules select smaller clusters for merging than 
%%smaller clusters first merge compared to  
%%than in 
the Achlioptas product rule (see Ref.~\cite{daCosta:ddgm10}).
%%\cite{Achlioptas:ads09}. 
%%Note that in 

In our rules, the selected nodes can belong to the same clusters. This happens frequently when a giant connected component is present in the network. Interestingly, if, in addition to rules (i), (ii), and (iii), we demand that 
%%%these events
the $2m$ nodes randomly chosen at each step must
%%of the model 
belong to different clusters (in this case, when samplings (i) and (ii) contain at least two nodes of $2m$ belonging to the same cluster, we reject these samplings and make new ones) 
 then 
%%%in the model, then 
%%instead of a single giant component, we will have three ones of the same size. 
there will be 
%%we will get 
not one 
but $2m-1$ giant connected components of the same size.
% On other words, if after stage (ii) we repeat stages (i) and (ii) when at least two of the $2m$ nodes belong to the same cluster, then $2m-1$ giant clusters will coexist with equal sizes in the percolation phase.
%, see Appendix~\ref{si0}. 
We observed this phenomenon in our simulations.
Note that multiple giant components were observed in simulations of other explosive percolation models, see for example Refs.~\cite{Chen:cs11,
%Nagler:ntg12,
Chen:cczcdn13,
Riordan:rw12}.
 In the rest of this paper we only consider the models defined by the rules of steps (i) to (iii), which do not demand that the $2m$ nodes considered at each step belong in different clusters.

%%%%%%%%%%%%%%%%%%%%%%%%%%%%%%%%%%%%%%%%%%%%%%%%%%%%%
%%%%%%%%%%%%%%%%%%%%%%%%%%%%%%%%%%%%%%%%%%%%%%%%%%%%%
%%%%%%%%%%%%%%%%%%%%%%%%%%%%%%%%%%%%%%%%%%%%%%%%%%%%%

\section{Evolution equations}
\label{s2}

%% 
%%%%%%%%%%%%%%%%%%%%%%%%%%%%%%%%%%%%%%%%%%%%%%%%%%
%%%%%%%%%%%%%%%%%%%%%%%%%%%%%%%%%%%%%%%%%%%%%%%%%%
\begin{figure*}[t]
%%%%[tbhd]
\begin{center}
\scalebox{0.41}{
\includegraphics[angle=0]{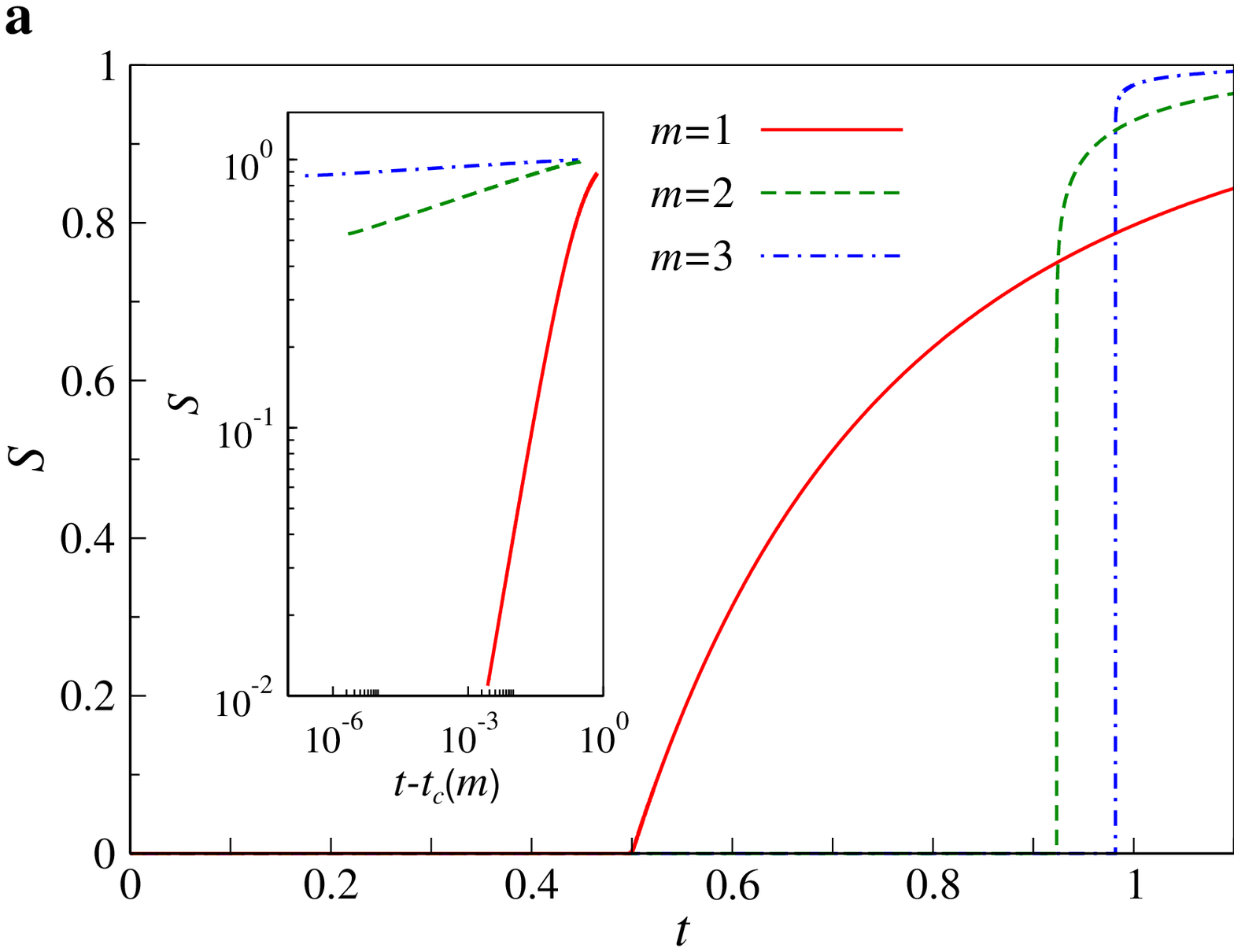}
\ \ \ \ \ \ \ \ \ \ \ \ 
\includegraphics[angle=0]{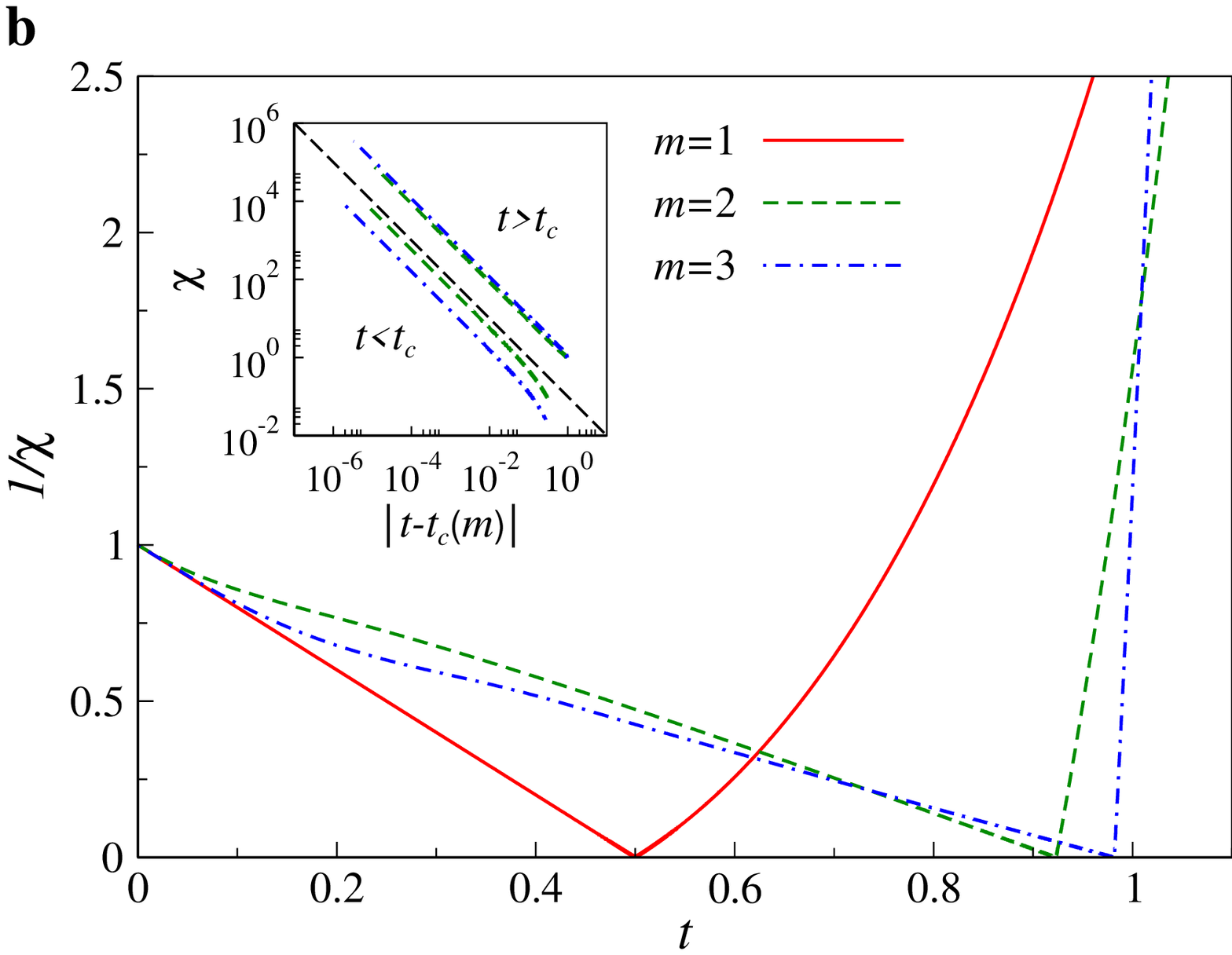}
}
\end{center}
\caption{ 
{\bf Relative size of the percolation cluster $S$ and the inverse susceptibility $1/\chi$ vs. $t$ for $m=1,2,3$.} 
Here and in Fig.~\ref{f3} the curves are the result of numerical solution of $10^5$ evolution equations. {\bf a}, despite the visually abrupt behavior of $S(t)$ at $m=2,3$, the inset suggests a power-law approach to the critical points. The slopes of the three curves in the inset are $1$, $0.0555$, and $0.0104$  for $m=1,2$, and $3$, respectively.  {\bf b}, the main panel  and the inset demonstrate the validity of the Curie--Weiss law for the susceptibility which is defined in the text. The black dashed guide line in the inset has slope $-1$. 
} 
\label{f2}
\end{figure*}
%%%%%%%%%%%%%%%%%%%%%%%%%%%%%%%%%%%%%%%%%%%%%%%%%%
%%%%%%%%%%%%%%%%%%%%%%%%%%%%%%%%%%%%%%%%%%%%%%%%%%

The evolution processes defined by these models can be treated as consecutive aggregation of clusters. 
%%, but the case of $m=1$ (ordinary percolation) is 
%%quite 
%%special. The difference is that while 
For standard percolation this process can be reversed (i.e., this is actually an equilibrium system), while for $m>1$ the process is irreversible.  
%%Thus the problem may be reduced to a specific aggregation process for which 
In order to describe this specific aggregation process we should find the 
%%analysis of the 
evolution of the size distribution $P(s)$ for a finite cluster of $s$ nodes to which a randomly chosen node belongs: 
$P(s)=sn(s)/\langle s\rangle$,  
where $n(s)$ is the size distribution of clusters (the probability that a uniformly randomly chosen cluster contains $s$ nodes), and $\langle s\rangle$ is the average 
%%cluster 
size for all clusters including the giant connected component. This distribution satisfies the sum rule $\sum_s P(s) = 1-S$. Here $S$ is the relative size of the percolation cluster. 
%%Note that for the sake of 
For brevity, we often do not indicate that the distributions are time dependent, where time $t$ is the ratio of the number of links and nodes in the system at a given step. 
%%Another basic characteristic in these processes is the size distribution of merging clusters selected by our procedure. 
We also introduce the probability $Q(s)$ that if we choose uniformly at random $m$ nodes then the smallest of the 
%%$m$ 
clusters to which these nodes belong 
is of size $s$. The sum rule here is $\sum_s Q(s) = 1-S^m$. The distribution $Q(s)$ can be easily expressed in terms of $P(s)$. Let us introduce the cumulative distributions 
%%
%%MULTIPLE GIANT COMPONENTS
%%
%%$P(s)$ 
%%
%%$P(s) = sn(s)/\langle s \rangle$
%%
%%$\sum_s P(s)=1-S$
%%
%%$m \geq 1$ 
%%
%%$Q(s)$ 
%%
%%$\sum_s Q(s)=1-S^m$ 
%%
$P_{\rm cum}(s) \equiv \sum_{u=s}^\infty P(u)$ 
and 
$Q_{\rm cum}(s) \equiv \sum_{u=s}^\infty Q(u)$, 
so that 
$P(s) = P_{\rm cum}(s) - P_{\rm cum}(s+1)$ 
and 
$Q(s) = Q_{\rm cum}(s) - Q_{\rm cum}(s+1)$.  
Then according to probability theory \cite{Feller:f68}, 
$Q_{\rm cum}(s)+S^m = [P_{\rm cum}(s)+S]^m$,   
%% 
%%$$Q_{\rm cum}(s)+S^m = [P_{\rm cum}(s)+S]^m. 
%%
%%$Q(s)=Q_{\rm cum}(s)-Q_{\rm cum}(s+1) = [P_{\rm cum}(s)+S]^m-[P_{\rm cum}(s+1)+S]^m$
%%
%%
%%
%%\begin{equation}
%%Q(s) = [P_{\rm cum}(s)
%%%%+S
%%+P_{\rm cum}(s+1)+2S]P(s) 
%%= [2-2P(1)
%%%%-2P(2)
%%-\ldots-2P(s-1)-P(s)]P(s)
%%, 
%%\label{e10}
%%\end{equation}
%%%%
%%
%%%%
%%\begin{equation}
%%Q(s) = P(s)\sum_{k=0}^{m-1}[P_{\rm cum}(s)+S]^k[P_{\rm cum}(s+1)+S]^{m-1-k}
%%, 
%%\label{e20}
%%\end{equation}
%%%%
%%
which gives 
\begin{align}
%Q(s) = P(s)\sum_{k=0}^{m-1}[1-\sum_{u=1}^{s-1}P(u)]^k[1-\sum_{u=1}^{s}P(u)]^{m-1-k}
Q(s) &= \left[1-\sum_{u=1}^{s-1}P(u)\right]^{m}\!\!-\ \left[1-\sum_{u=1}^{s}P(u)\right]^{m}
%%\label{e50}
\nonumber
\\
&=\sum_{k=1}^{m}\binom{m}{k}P(s)^k\left[1-\sum_{u=1}^{s}P(u)\right]^{m-k}
, 
\label{e100}
\end{align}
that is, $Q(s)$ is determined by $P(s')$ with $s'\leq s$. 
%%Here for the sake of brevity we  
For the infinite system, the evolution of these coupled distributions are described exactly by the the infinite set of evolution 
%%master (rate) 
equations: 
\begin{equation}
\frac{\partial P(s,t)}{\partial t}
= s \sum_{u+v=s} Q(u,t)Q(v,t) - 2 sQ(s,t)
. 
\label{e200}
\end{equation}
We derived these equations in a similar way to ordinary percolation \cite{Leyvraz:l03,Krapivsky:krb10}. 
This is actually a version of Smoluchowski equation \cite{Smoluchowski:s1915} for our aggregation process. 
%%We derived it in a similar way to ordinary percolation \cite{Leyvraz:l03,Krapivsky:krb10}. 
For ordinary percolation, $Q(s,t)=P(s,t)$, and the problem is explicitly solvable  \cite{Leyvraz:l03,Krapivsky:krb10}. 
%%The only difference from the evolution equations for ordinary percolation (specifically, the classical random graph model, which corresponds to $m=1$) is that the distribution $Q(s,t)$ on the right-hand side of equation~(\ref{e200}) should be substituted by $P(s,t)$ for ordinary percolation. 
%%One can see that for 
For $m>1$, the right-hand side of Eq.~(\ref{e200}) is not bilinear, and the explicit solution is not possible.  
%%is a difficult nonlinear equation which cannot be solved by using generating functions in contrast to ordinary percolation. 
We have solved this system of equations numerically for $s\leq 10^6$ in the case of $m=2$ \cite{daCosta:ddgm10}. 
Figure~\ref{f2}a shows the dependence of the relative size of the percolation cluster obtained in this way for $m=1,2$, and $3$. In the following we present an exact solution of the problem in the critical region around the percolation threshold $t_c$, where the distributions have scaling form. 
We will first assume that the transition is continuous, then derive equations for the scaling functions of $Q(s,t)$ and $P(s,t)$. By solving these equations, we will demonstrate that the scaling functions exist and that our assumption is correct and self-consistent. 
%%%%So here we solve equation~(\ref{e200}) for clusters of all sizes without seaching for the value of $t_c$ which is dependent on an initial cluster size distribution and so is not of great interest for us.   

%%Summing both sides of equation~(\ref{e200}) over $s$, we obtain  
%%
%%eq for $S$
%%
%%which leads to the delay
%%
%%eq for the first moment 

Equation~(\ref{e200}) leads to the following equations for the moments of the distributions and the size of the percolation cluster: 
\begin{eqnarray}
\frac{\partial S}{\partial t} 
& = &
2 S^m \langle s \rangle_{\scriptscriptstyle \!Q} 
,
%%.
\label{e300}
\\[5pt]
%%\end{equation}
%%
%%and
%%
%%\begin{equation}
\frac{\partial \langle s \rangle_{\scriptscriptstyle \!P}}{\partial t}
& = &
2 \langle s \rangle^2_{\scriptscriptstyle \!Q} - 
2 S^m \langle s^2 \rangle_{\scriptscriptstyle \!Q} 
, 
%%,
\label{e400}
\end{eqnarray}
where $\langle s^n  \rangle_{\scriptscriptstyle \!P}=\sum_s s^nP(s)$ and $\langle s^n \rangle_{\scriptscriptstyle \!Q}=\sum_s s^nQ(s)$. We can use these equations to derive relations between critical exponents.
%%%%Equation~(3) demonstrates the principal difference of ``explosive'' percolation from ordinary one. Let us seed a giant component of relative size $h\ll 1$ in the normal phase at some moment $t<t_c$ 
%%, that is in the normal phase, 
%%%%and consider its 
%%immediate 
%%%%evolution. Equation~(3) shows that the growth rate of this component is proportional to $h^m$, i.e., it is severely suppressed in the entire normal phase if $m>1$. 
%%Only for ordinary percolation, the corresponding growth rate is proportional to the size $h$ of the seed component. 
%%%%This supression results in the delayed transition compared to $m=1$. Equations~(\ref{e300}) and (\ref{e400}) will allow us to obtain conveniently relations between critical exponents. 
Their interpretation is given in Appendix~\ref{si1}. 

%%%%%%%%%%%%%%%%%%%%%%%%%%%%%%%%%%%%%%%%%%%%%%%%%%%%
%%%%%%%%%%%%%%%%%%%%%%%%%%%%%%%%%%%%%%%%%%%%%%%%%%%%
%%%%%%%%%%%%%%%%%%%%%%%%%%%%%%%%%%%%%%%%%%%%%%%%%%%%

\section{Basic scaling relations}
\label{s3}
%%%%%%%%%%%%%%%%%%%%%%%%%%%%%%%%%%%%%%%%%%%%%%% my version
In this section we find the basic scaling relations for the explosive percolation models.
We assume that in the critical region (both below and above $t_c$) the distribution function $P(s,t)$ for large $s$  has a  scaling form,
\begin{equation}
P(s,t) = s^{1-\tau}f(s\delta^{1/\sigma})
,
\label{e500}
\end{equation}
where $\delta=|t-t_c| \ll 1$, $\tau$ and $\sigma$ are critical exponents, and $f(x)$ is a scaling function. Substituting Eq.~(\ref{e500}) into the sum rule $\sum_s P(s,t) = 1-S$ at $t\geq t_c$  and using the equality $\sum_s P(s,t_c) = 1$ at $t= t_c$,
we find the size of the giant component, 
\begin{equation}
S = \sum_{s} [P(s,t_c)-P(s,t)] \propto \delta^{\beta},
\label{e700}
\end{equation}
where the critical exponent $\beta$ is 
\begin{equation}
\beta = (\tau-2)/\sigma , 
\label{e600}
\end{equation}
see Appendix~\ref{si3new}
%Information 
for detailed derivation and discussion.
%%This result is obtained by assuming that $x^{2-\tau}[f(0){-}f(x)] {\to} 0$ at $ x {\to} 0$. Moreover, the integral $\int_{0}^\infty  x^{2-\tau} [df(x)/dx] dx$ must be finite. These assumptions impose conditions on the value of $\tau$ and the behavior of $f(x)$ at $x \ll 1$ and $x \gg 1$ \cite{Stauffer:s79}. In particular, we demand $2 < \tau < 2+\sigma$ at $m \geq 2$ in contrast to $2 < \tau < 3$ at $m=1$ \cite{Stauffer:s79}. Our solution presented for $m \geq 2$ in Sec. \ref{si5} (Supplementary Information) satisfies these conditions.

The scaling form of the distribution $Q(s,t)$ in the normal phase of the transition is found by substituting Eq.~(\ref{e500}) into Eq.~(\ref{e100}). Using the fact that
\begin{equation}
Q(s,\delta)  \cong  m\Bigl(\int_s^\infty du P(u,\delta)\Bigr)^{m-1} P(s,\delta)
,
\label{e800}
\end{equation}
at large $s$, we obtain 
\begin{equation}
Q(s,\delta) = s^{(2m-1)-m\tau}g(s\delta^{1/\sigma})
,
\label{e900}
\end{equation}
where $g(x)$ is a scaling function related with $f(x)$.

The critical behavior of the first moments of the distributions, $\langle s \rangle_P 
=\sum_s s P(s)
\sim \delta^{-\gamma_P}$ 
and $\langle s \rangle_Q 
=\sum_s s Q(s)
\sim \delta^{-\gamma_Q}$, 
easily follows from Eqs.~(\ref{e500}) and (\ref{e900}). We  find 
\begin{eqnarray}
\gamma_P &=&
(3-\tau)/\sigma,
\label{e1000}
\\[5pt]
\gamma_Q &=&
(2m+1-m\tau)/\sigma.
\label{e1100}
\end{eqnarray}
From Eq.~(\ref{e400}), we obtain the relation $\gamma_P + 1 = 2\gamma_Q$ which allows us express all the critical exponents in terms of a single unknown exponent, for example, $\beta$:
\begin{eqnarray}
\tau & = & 2 + \frac{\beta}{1 + (2m-1)\beta}
,
\label{e1400}
\\[5pt]
1/\sigma & = & 1 + (2m-1)\beta
,
\label{e1500}
\\[5pt]
\gamma_P & = & 1 + 2(m-1)\beta
,
\label{e1600}
\\[5pt]
\gamma_Q & = & 1 + (m-1)\beta
.
\label{e1700}
\end{eqnarray}
%%

%%%%%%%%%%%%%%%%%%%%%%%%%%%%%%%%%%%%%%%%%%%%%%%%%%
%%%%%%%%%%%%%%%%%%%%%%%%%%%%%%%%%%%%%%%%%%%%%%%%%%
%%%%%%%%%%%%%%%%%%%%%%%%%%%%%%%%%%%%%%%%%%%%%%%%%%

\section{ 
%%Nature of the 
Order parameter and susceptibility}
\label{s4}

%%The divergence of the susceptibility manifests/signals the critical point of the explosive percolation transition. 

%%%%%%%%%%%%%%%%%%%%%%%%%%%%%%%%%%%%%%%%%%%%%%%%%%%
%%%%%%%%%%%%%%%%%%%%%%%%%%%%%%%%%%%%%%%%%%%%%%%%%%%
\begin{figure}[]
%%%%[tbhd]
\begin{center}
\scalebox{0.415}{\includegraphics[angle=0]{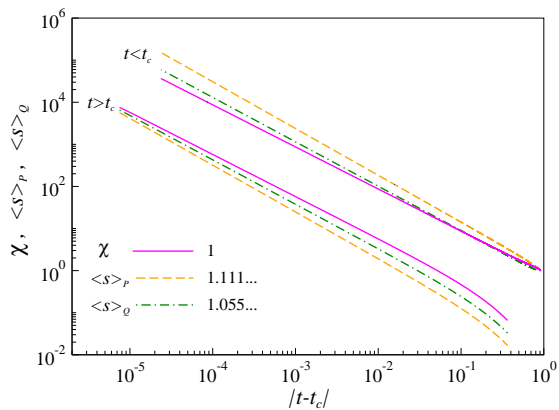}}
\end{center}
\caption{ 
{\bf Comparison between susceptibility and the first moments of the distributions $P(s)$ and $Q(s)$ vs. deviation from the critical point for $m=2$.} The numbers show the slopes of the correspondent curves. 
The susceptibility demonstrates the Curie--Weiss law, while the critical exponents of $\langle s \rangle_P$ and $\langle s \rangle_Q$ agree with relations (\ref{e1600}) and (\ref{e1700}) when $\beta=0.055...\,$. 
} 
\label{f3}
\end{figure}
%%%%%%%%%%%%%%%%%%%%%%%%%%%%%%%%%%%%%%%%%%%%%%%%%%%
%%%%%%%%%%%%%%%%%%%%%%%%%%%%%%%%%%%%%%%%%%%%%%%%%%%

%I is not true that Òan order parameter could be anything that distinguish the two phases (zero above and nonzero below some critical point).Ó 
For continuous phase transitions the order parameter cannot be chosen in an arbitrary way, by demanding only that it is zero in the normal phase and non-zero in the ordered phase.
For these transitions the order parameter must satisfy several strict conditions that are well known in the theory of phase transitions~\cite{Stanley_s-book-71}.
 First, the critical exponent of the order parameter must satisfy basic hyperscaling relations discussed in Sec.~\ref{s5} and Appendix~\ref{si3}. Second, the order parameter, susceptibility, and pair correlation function are closely related to each other. 
 Indeed, the susceptibility is the derivative of the order parameter with respect to a conjugate field.
From this, the relation between susceptibility and pair correlation function follows, which determines a basic relation between the critical exponents of these physical quantities. 
%In particular, the susceptibility must diverge in the critical point. 
In the case of ordinary percolation, the relations between the order parameter, susceptibility, and pair correlation function were obtained rigorously by use of the one-state limit of the Potts model~\cite{Kasteleyn:kf69}.
%(see Wu, Rev. Mod. Phys. (1982)).
 %It is important to outline
We stress that the order parameter, the susceptibility, and the pair correlation function found here for explosive percolation satisfy all of these basic conditions.
 %The history of 
 In statistical and solid state physics there are many examples (such as spin glasses~\cite{Mezard:mpv-book-86}, percolation~\cite{Stauffer:sa-book94}, etc)
 demonstrating that the search for
%%  where it is the necessity to satisfy the fundamental physical relations between the order parameter, the susceptibility, and the pair correlation function that makes the problem of choosing 
the order parameter 
%%to be very nontrivial.
is a nontrivial problem.

For the ordinary percolation phase transition, the relative size of the percolation cluster $S$ is the order parameter, while the average size $\langle s \rangle_P$ of a finite cluster, to which a uniformly randomly chosen node belongs, 
plays the role of susceptibility. 
%%%%the average size $\langle s \rangle_P$ of a finite cluster to which a uniformly randomly chosen node belongs , and the relative size of the percolation cluster $S$ (i.e., the probability that a uniformly randomly chosen node is in the percolation cluster), play the role of susceptibility and the order parameter, respectively. 
%%for ordinary percolation. 
So for ordinary percolation, the exponents 
$\beta$ and $\gamma_P$ are the critical exponents of the order parameter and susceptibility. 
%%$\gamma_P$ and $\beta$ are the critical exponents of susceptibility and the order parameter. 
%%, for which exponent $\gamma\equiv\gamma_P$ equals $1$ for classical random graphs. 
%%%%in systems of dimensionality higher than the upper critical dimension. 
%%This is not the case for the explosive percolation models. 
Let us show that the susceptibility and the order parameter for explosive percolation have a quite different meaning. Here we present heuristic arguments,  
%%that the order parameter and susceptibility for explosive percolation principally differ from ordinary percolation, 
%%%%. have a quite different meaning., 
for a 
%%more strict derivation 
comprehensive consideration see Appendix~\ref{si2}. 

%%Susceptibility in cooperative models of statistical mechanics is expressed in terms of the average value of the pairwise spin-spin correlator, where averaging is over all spins of a system. 
%%For percolation problems which we consider, this is the probability $c_2$ that a new link interconnects nodes in the same cluster \cite{}. 
For percolation problems which we consider, the probability $c_2$ that 
a new link interconnects nodes in the same cluster
%%two nodes chosen using the rules of the corresponding model belong to the same cluster, i.e. the new connection is within one cluster. 
%%This probability 
provides both susceptibility $\chi$ and the order parameter $\phi$ of the system: $c_2=\chi/N + \phi^2$, where the second summand is the probability that both nodes belong to the percolation cluster \cite{Stauffer:s79,Stauffer:sa-book94}. 
To measure the susceptibility in this process experimentally, one should find the fraction of events in which two nodes selected by the specific rule of the model fall into the same finite cluster. The divergence of the susceptibility manifests the critical point of the explosive percolation transition.

For our model of explosive percolation, the probability that two nodes selected by 
%%using the algorithm from the previous section 
our algorithm belong to the same cluster is 
\begin{equation}   
c_2 = 
%%\frac{\langle s \rangle}{N} 
%%\sum_s \frac{Q^2(s)}{n(s)N/\langle s \rangle} + 
\frac{1}{N}\sum_s \frac{sQ^2(s)}{P(s)} +
%%\frac{S^{2m}}{1}
S^{2m}
.
\label{e1800}
\end{equation}
For rigorous derivation of this expression, see Appendix~\ref{si2}. 
The first term on the right-hand side is the probability that both selected nodes belong to the same finite cluster, while the second term is the probability that both selected nodes are in the 
%%belong to the 
%%giant connected component, 
percolation cluster.  
To obtain the first term, we divide  
%%$Q^2(s)$ is 
the probability $Q^2(s)$ that both selected nodes belong to clusters of size $s$ by 
%%and $n(s)N/\langle s \rangle = P(s)N/s$ is 
the number of clusters of $s$ nodes in the system, $n(s)N/\langle s \rangle = P(s)N/s$, and then sum over $s$. The first term gives the susceptibility for the explosive percolation model (divided by $N$), the second term gives the square of the order parameter. Consequently the order parameter in these models is $S^m$ and not $S$ as is usually believed. 
%%which is $M=S^m$. 
In particular, at $m=1$, Eq.~(\ref{e1800}) is reduced to the well-known relation $c_2 = \langle s \rangle_P/N + S^2$ for ordinary percolation. Substituting the scaling forms of the distributions $P(s,t)$ 
%%=sn(s,t)/\langle s \rangle$ 
and $Q(s,t)$ near the critical point, Eqs.~(\ref{e500}) and (\ref{e900}), respectively, into Eq.~(\ref{e1800}) immediately gives $\chi \sim \delta^{-\gamma}$, where the critical exponent of susceptibility is $\gamma=1$. This is the Curie-Weiss law which is valid for 
%%the critical singularity of susceptibility in 
cooperative systems above an upper critical dimension, where mean-field theories work. 
The inset of Fig.~\ref{f2}b confirms this law for $m=2,$ and $3$. Notice in Fig.~\ref{f2}b that while for ordinary percolation ($m=1$), the moduli of the slopes of $1/\chi(t)$ 
above and below the transition are equal as $t\to t_c$,
%approaching the critical point from above and below the transition are equal, 
for higher $m$ they differ drastically from each other.
Figure~\ref{f3} demonstrates the contrast between the critical divergencies of the susceptibility (which diverges according to the Curie-Weiss law) and the first moments $\langle s \rangle_P$ and $\langle s \rangle_Q$ for $m=2$. 
%%compares critical divergencies of the susceptibility with the first moments $\langle s \rangle_P$ and $\langle s \rangle_Q$ for $m=2$ and indicates their critical exponents. The figure demonstrates the susceptibility that s

%%$$
%%\gamma_Q = (m+1-m\tilde{\tau})/\sigma = (2m+1-m\tau)/\sigma
%%$$
%%
%%$$
%%\gamma_P + 1 = 2\gamma_Q
%%$$
%%
%%So 
%%$$
%%\sigma = 2m - (2m-1)\tilde{\tau}
%%$$
%%
%%$$
%%1/\sigma = 1 + (2m-1)\beta
%%$$
%%
%%$$
%%\gamma_P = 1 + 2(m-1)\beta
%%$$
%%
%%$$
%%\gamma_Q = 1 + (m-1)\beta
%%$$
%%
%%$$
%%\tau = 2 + \frac{\beta}{1 + (2m-1)\beta}
%%$$
%%
%%$$
%%\beta = \frac{1}{2m-1}\,\frac{\tilde{\tau}-1}{2m/(2m-1)-\tilde{\tau}}
%%$$
%%
%%
%%$\tau$  
%%
%%$\sigma$
%%
%%$\beta$ 
%%
%%$\gamma_P$ 
%%
%%$\gamma_Q$ 
%%
%%$\nu=1/2$, $\eta=0$
%%
%%$$
%%p_2 = \frac{\langle s \rangle}{N} \sum_s \frac{Q^2(s)}{n(s)} + \frac{S^{2m}}{1}
%%$$
%%
%%The first term - susceptibility, $S^m$ is the order parameter. 

%%%%%%%%%%%%%%%%%%%%%%%%%%%%%%%%%%%%%%%%%%%%%%%%
%%%%%%%%%%%%%%%%%%%%%%%%%%%%%%%%%%%%%%%%%%%%%%%%
%%%%%%%%%%%%%%%%%%%%%%%%%%%%%%%%%%%%%%%%%%%%%%%%

\section{Hyperscaling}
\label{s5}

Another set of relations between critical exponents contain the dimensionality $d$ of a system, the fractal dimension $d_f$ of clusters at the critical point, the correlation length critical exponent $\nu$, and the Fisher exponent $\eta$ \cite{Privman:p90}. These relations are often called hyperscaling relations. In this work we consider infinite-dimensional models, but they can be generalized and formulated for an arbitrary $d$. For this generalization, one can 
%%easily 
derive the hyperscaling relations for $d$ below the upper critical dimension $d_u$ in the same way as for ordinary percolation (see Appendix~\ref{si3}). The resulting hyperscaling relations at $d<d_u$ are as follows:  
\begin{eqnarray}
d_f & = & 1/(\sigma\nu)
,
\label{e1900}
\\[5pt]
d_f & = & d-\beta/\nu
,
\label{e2000}
\\[5pt]
d-2+\eta & = & 2\beta^*
%%_{{\rm order\ parameter}}
\!/\nu
, 
\label{e2100}
\end{eqnarray}
where the order parameter critical exponent $\beta^*
%%_{{\rm order\ parameter}}
$ equals $m\beta$ for our models, since the order parameter is $S^m$.  
%%as was shown above. 
%%One should stress that the exponent $\beta$ of the percolation cluster enters into relation (\ref{e2000}) in contrast to equation~(\ref{e2100}) with $\beta^*
%%_{{\rm order\ parameter}}
%%$. 
%%At $m=1$, equations~(\ref{e1900})--(\ref{e2100}) are reduced to the standard hyperscaling relations for ordinary percolatin. 
%%To obtain the hyperscaling relations a

Above the upper critical dimension, one should set in Eqs.~(\ref{e1900})--(\ref{e2100}) the exponents $\nu$ and $\eta$ to their mean-field theory values, $1/2$ and $0$, respectively, and $d$ to $d_u$ \cite{Privman:p90}. The resulting relations together with Eq.~(\ref{e1500}) lead to the following expressions for the fractal and upper critical dimensions $d_f$ and $d_u$ in terms of the exponent $\beta$: 
%%
%%Above the upper critical dimension, $d\to d_u$, $\nu\to 1/2$, $\eta\to 0$ \cite{Privman:p90}. 
%%
%%Plus note that $\beta_{order\ param}=m\beta$. 
%%
\begin{eqnarray}
d_f & = & 2[1 + (2m-1)\beta]
,
\label{e2200}
\\[5pt]
d_u & = & 2 + 4m\beta
. 
\label{e2300}
\end{eqnarray}
These relations demonstrate that both $d_f$ and $d_u$ are very close to $2$ when $m>1$. This means that explosive percolation models of this kind defined in two dimensions have critical features very similar to those predicted by the mean-field theory.

%%%%%%%%%%%%%%%%%%%%%%%%%%%%%%%%%%%%%%%%%%%%%%%%%%%
%%%%%%%%%%%%%%%%%%%%%%%%%%%%%%%%%%%%%%%%%%%%%%%%%%%
%%%%%%%%%%%%%%%%%%%%%%%%%%%%%%%%%%%%%%%%%%%%%%%%%%%

\section{Scaling functions and exponents} 
\label{s6}
%%{Equations for scaling functions} 

In this section we outline the derivation of the equations for scaling functions and describe their solutions. 
Above the percolation threshold,  Eq.~(\ref{e100}) is reduced to $Q(s) \cong mS^{m-1}P(s)$   at large $s$, which makes the resulting evolution equation for $P(s,t)$ to be similar to that for ordinary percolation.  
%%and so easily solvable with the initial condition $P(s,t_c) \cong A s^{-\tau+1}$. 
In our work \cite{daCosta:ddgm10} we assumed that the distribution $P(s)$ at the critical point is, asymptotically, a power law,  %%we can easily 
which enabled us to solve Eq.~(\ref{e200}) using the initial condition $P(s,t_c) \cong A s^{1-\tau}$ (in \cite{daCosta:ddgm10} we show the solution for $m=2$, for an arbitrary $m$ see Appendix~\ref{si9}). This provides the scaling functions $f(x)$ and $g(x)$ on the upper side of the phase transition in terms of the yet unknown critical amplitude $A$ and exponent $\tau$ \cite{daCosta:ddgm10}. 
%%We are interested not ....
%%In our previous work \cite{xxx}, we have shown that if it is known that the distribution $P(s)$ at the critical point is, asymptotically, power-law with some given critical exponent and amplitude, $P(s,t_c) \cong A s^{-\tau+1}$, as it should be for a continuous phase transition, then from equation~(\ref{e200}), immediately follows the power law $S \cong B \delta^\beta$, where 
%%$\beta = (\tau-2)/[1 - (2m-1)(\tau-2)]$ 
%%as in equation~(\ref{e1400}) and the coefficient $B$ is expressed in terms of $A$ and $\tau$. 
%%(Here the critical amplitude $A$ is determined by the initial form of the distribution $P(s,t=0)$.)
%%Furthermore, this assumption allows us to find the scaling functions $f(x)$ and $g(x)$ on the upper side of the phase transition, i.e. at $t>t_c$. The form of these functions turn out to be close to exponential, similarly to ordinary percolation above an upper critical dimension. That derivation exploited the convenient simplification of the equations above the critical point, where $S$ differs from zero. In this region, at large $s$, equation~(\ref{e100}) is reduced asymptotically to $Q(s) \cong mS^{m-1}P(s)$, which makes the resulting evolution equation for $P(s,t)$ to be similar to that for ordinary percolation and so easily solvable with the initial condition $P(s,t_c) \cong A s^{-\tau+1}$. 
In the present work we will 
%%Consequently our present 
%%more difficult 
%%task is to 
obtain the distribution at the critical point 
%%, used in that derivation, 
and verify its power-law form. We will find the critical exponent and amplitude of $P(s,t_c)$, 
%%its critical exponent (if this distribution will appear to be power-law), 
and obtain the scaling functions 
%%on the normal phase side of the phase 
below the transition. 
%%, i.e. at $t<t_c$. So, simultaneously we verify that the transition is continuous.  
In this way we will completely describe the cluster size distribution in the entire critical region.

Let us approach the critical point from the normal-phase side. 
To derive equations for scaling functions, we have to remove the non-scaling, low $s$ parts of the distributions from Eqs.~(\ref{e100}) and~(\ref{e200}) and then substitute their scaling forms of Eqs.~(\ref{e500}) and~(\ref{e900}). 
As a result we arrive at a system of nonlinear integro-differential equations of the second order, convenient for analytical and numerical treatment. 
The details of the derivation and the resulting equations for the scaling functions are presented in Appendix~\ref{si4}. 
In essence, these are nonlinear eigenfunction equations, where eigenfunctions are the scaling functions of our problem and the eigenvalue is one of the critical exponents, e.g., $\tau$.  
These equations are solved on the one-dimensional interval $0\leq x<\infty$.   
At $x=0$, $f(x)$ and $g(x)$ coincide  with the critical amplitudes for the corresponding distributions: 
$P(s,t_c) \cong f(0) s^{1-\tau}$ and $Q(s,t_c) \cong g(0)s^{(2m-1)-m\tau}$, respectively. 
%%The boundary conditions are the following. 
%%%%%As the boundary condition at infinity we demand decay of the scaling functions to zero, staying positive. 
The amplitude $g(0)$ can be expressed in terms of $f(0)$ and $\tau$. 
%%In its turn, $f(0)$ depends on initial conditions, that is on $P(s,t=0)$. On the other hand, $\tau$ is independent of initial conditions. 
%%
The critical amplitude $f(0)$, as well as the detailed shapes of the scaling functions, is determined by the initial distribution of cluster sizes, 
%%at $t=0$, 
$P(s,t=0)$. In contrast to that, the critical exponents  
%%values 
do not depend on 
%%the 
initial conditions. 
%%, if $P(s,t=0)$ decays sufficiently rapidly (see below). 
%%(Note however that, unlike the critical exponents, the scaling function shapes depend on the initial conditions.)   
So, 
%%while 
when searching for the solution of the equation,  
%%Since the resulting value of the critical exponent $\tau$ is the same for a wide range of initial conditions (the distribution $P(s,t=0)$ must decay faster than ... finite susceptibility ), and we are interested only in $\tau$, 
we can set any convenient value of the critical amplitude $f(0)$. 
%%to find the solution of the equation. 
For different values of $f(0)$, the resulting value of the critical exponent $\tau$ should be the same, and the scaling functions, while differing from each other, should be qualitatively 
%%very 
similar. 
For a given critical amplitude $f(0)$, the system of first order differential equations for the scaling functions shown in Eq.~(\ref{se3300}) can be 
%%easily 
directly solved numerically. This solution gives the exponent $\tau$ together with the scaling functions $f(x)$ and $g(x)$. 
%%for a given $f(0)$. 
%%To .....   condition
The unknown critical exponent $\tau$ is obtained from the condition that $f(x)$ and $g(x)$ decay to zero as $x\to\infty$, 
%%at infinity, 
while staying positive 
%%as $x$ approaches infinity 
(see Appendix~\ref{si7} for details of the numerical procedure). 
These calculations converge rapidly giving the final value of $\tau$ and the scaling functions with any desired precision, i.e. exactly in a physics sense. 

%%%%%%%%%%%%%%%%%%%%%%%%%%%%%%%%%%%%%%%%%%%%%%%%%%%%
%%%%%%%%%%%%%%%%%%%%%%%%%%%%%%%%%%%%%%%%%%%%%%%%%%%%
\begin{figure}[t]
%%%%[tbhd]
\begin{center}
\scalebox{0.415}{\includegraphics[angle=0]{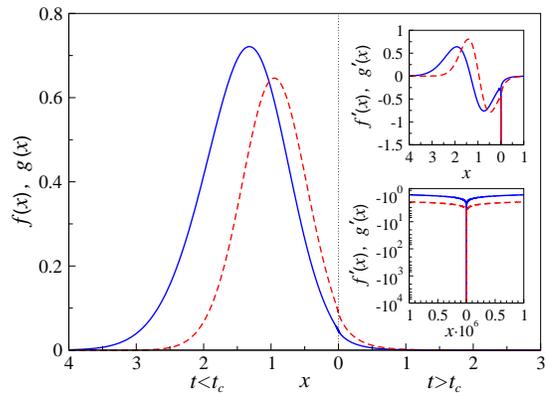}}
\end{center}
\caption{ 
\textbf{Scaling functions $f(x)$ and $g(x)$ for $m=2$.} The solid lines are for $f(x)$, and the dashed lines are for $g(x)$. The insets showing the respective derivatives highlight the presence of singularities at $x=0$. 
} 
\label{f4}
\end{figure}
%%%%%%%%%%%%%%%%%%%%%%%%%%%%%%%%%%%%%%%%%%%%%%%%%%%%
%%%%%%%%%%%%%%%%%%%%%%%%%%%%%%%%%%%%%%%%%%%%%%%%%%%%

%%\section*{Scaling functions and exponents} 

The resulting scaling functions $f(x)$ and $g(x)$ are shown in Fig.~\ref{f4} for $m=2$ (for higher $m$ the scaling functions are qualitatively similar). The plot shows scaling functions 
%%for 
in the normal phase, $t<t_c$,  
%%(the left-hand side) 
and 
%%for 
in the phase with the percolation cluster, $t>t_c$. 
%%(the right-hand side). 
%%The normal phase side was obtained from the scaling function equations in the way described above. 
%%The resulting asymptotics $P(s,t_c) \cong f(0) s^{1-\tau}$ was used as the initial condition for  
%%This result also provided us with an initial condition at $t_c$ for 
%%the linearized evolution equations in the phase with a giant component. Their solution gave the right-hand side of the plot. 
This figure demonstrates a drastic contrast with the scaling function for ordinary percolation above the upper critical dimension, which is symmetric. It is the exponential function  $f(x) = e^{-2x}/\sqrt{2\pi}$ 
%%and exponentially decaying 
both for $t>t_c$ and $t<t_c$. 
Note that a similar asymmetry is observed at 
%%dimensions 
$1<d<d_u=6$ in ordinary percolation~\cite{Nakanish:ns80}. 
%%Besides, at low $d$, the exponent $\tau$ of ordinary percolation is also close to $2$ (Table~\ref{t:perc_f(0)_exponents}). At $d=2$ for example, $\tau \simeq 2.139$. 
%%In explosive percolation models, the asymmetry of scaling functions persists even in the limit $d\to\infty$. 
%%
%2 singularities at the 0 
The insets in Fig.~\ref{f4} demonstrate that the scaling functions have singularities at $x=0$. 
%%The top inset shows the derivatives of the curves plotted in the main panel for the same range of $x$, and the bottom one shows a zoom of the near $x=0$ region. The derivatives of $f(x)$ and $g(x)$ diverge approaching $x=0$, from both sides. This singular behavior is correctly predicted by equations~(\ref{eq:Taylor_f_g}), that were derived for the disordered phase, but, in fact, hold on both sides of the transition. It is clear that $f(0)$ and $g(0)$ should be equal on both phases. Moreover, the series expansion coefficients of the singularities of $f(x)$ and $g(x)$ above $t_c$, can be found in a similar way that for $t<t_c$, showing that, for the ordered phase, the coefficients $a_1,b_1,a_2,b_2,\dots$ are simply the inverse of the ones for the disordered phase (i.e, in the percolation phase, $-a_1,-b_1,-a_2,-b_2,\dots$ are also given expressions~(\ref{eq:a1_b1_a2_b2_SEPQ})).
% are also given by expressions~\ref{eq:a1_b1_SEPQ}
In Appendix~\ref{si5} we show that $f(x)-f(0)  \propto g(x)-g(0) \propto x^\sigma$ near $x=0$ at $m>1$, where the critical exponent $\sigma$ is slightly smaller than $1$, see Eq.~(\ref{e1500}). 
%m 
%%We have solved numerically the exact system of differential equations~(\ref{eq:sol_diff_system}), for the scaling functions below the percolation threshold, using expansions~(\ref{eq:Taylor_f_g}) as initial conditions. 
%%With the method developed above we could calculate the value of 
%%We found $\tau$ for $m=2,3,\dots,20$. 
%%The ordinary differential equations can be solved with any desired precision, therefore, the corresponding eigenfunction, $\tilde{f}(x)$, and eigenvalue, $\tau$, may be obtained with any precision as well. 
%%The scaling functions stay qualitatively similar with increasing $m \geq 2$ (for $m=2$ they are shown in Figure~\ref{f:scal_func_SEPQ}). They show a maximum at some $x>0$ is the region of $t<t_c$, and decay monotonically in the region of $t>t_c$. 
%%The main difference between models is the asymptotic behavior of the functions on the $t<t_c$ phase. 
Below $t_c$ for large $x$ we find $f(x)\propto \exp\left(-Cx^{1+\ln m/\ln 2}\right)$ and $g(x)\propto \exp\left(-mCx^{1+\ln m/\ln 2}\right)$, where $C$ is a constant (see Appendix~\ref{si6}). Above $t_c$ the scaling functions exponentially decay to $0$. 
%% (see expressions~(\ref{eq:f_g_m_asymp})). 
%%On the other hand, the value of $\tau$ indeed approaches $2$ quite rapidly with growing $m$, as the results of section~\ref{CEPT} previously suggested. This section's results are summarized in Table~\ref{t:SEPQ}. 
%While the scaling functions stay qualitatively similar with increasing $m \geq 2$ (for $m=2$ they are shown in Figure~\ref{f:scal_func_SEPQ}), the value of $\tau$ indeed approaches $2$ quite rapidly with $m$, as previously suggested by the results of section~\ref{CEPT}. This section's results are summarized in Table~\ref{t:SEPQ}. 
%%
%%%%%%%%%%%%%%%%%%%%%%%%%%%%%%%%%%%%%%%%%%%%%%%%%%
%%%%%%%%%%%%%%%%%%%%%%%%%%%%%%%%%%%%%%%%%%%%%%%%%%
\begin{figure}[t]
%%%%[tbhd]
\begin{center}
\scalebox{0.43}{\includegraphics[angle=0]
{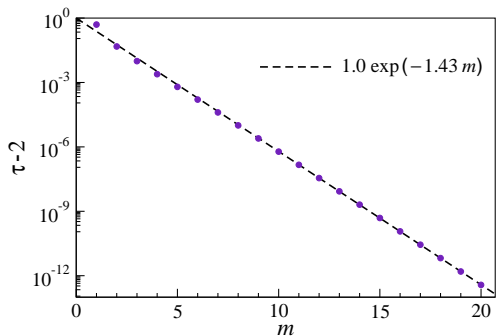}
%%{sol_expl_perc_quest-tau-2_vs_m2.eps}
}
\end{center}
\caption{ 
\textbf{Deviation of the critical exponent $\tau$ from $2$ vs. $m$.} The dashed line (exponential function) fits well the dots obtained by our numerical procedure. For the corresponding numerical values see Appendix~\ref{si8}. 
} 
\label{f5}
\end{figure}
%%%%%%%%%%%%%%%%%%%%%%%%%%%%%%%%%%%%%%%%%%%%%%%%%%
%%%%%%%%%%%%%%%%%%%%%%%%%%%%%%%%%%%%%%%%%%%%%%%%%%

We found $\tau$ for $m$ up to $20$, see Fig.~\ref{f5} and Table~\ref{st1}, which shows that $\tau-2$ (as well as $\beta$) decreases with $m$ as $\exp(-1.43 m)$. All other exponents and upper critical dimensions can be readily found from the relations between critical exponents. In particular, the upper critical dimension decreases rapidly to $2$ with increasing $m$.

\section{Discussion}
\label{s7}

Our theory reveals that the explosive percolation problem is a direct generalization of ordinary percolation, but this generalization turned out to be principally non-trivial. The complete scaling description which we developed explains the genuine continuous nature of the explosive percolation phase transitions in the investigated class of systems. 
We completely described all scaling properties of this transition within the framework of the theory of continuous phase transitions.  
So a complete description of the class of processes considered in this work does not require the introduction of new notions, like ``weakly discontinuous transition''.
%%So there is no need to introduce new notions, like ``weakly discontinuous transition''.  
%%\cite{Nagler:nlt11}. 
Despite of this continuity, we found and highlighted the drastic difference between ordinary and explosive percolation. 
%%%We found that, unlike in ordinary percolation, the order parameter and susceptibility of explosive percolation are not 
%%$S$ 
%%%the relative size of the percolation cluster and 
%%$\langle s \rangle_P$. 
%%%the average size of a finite cluster to which a random node belongs. 
%%in contrast to 
%%differ from those for 
%%ordinary percolation. 
We found the order parameter and susceptibility of explosive percolation and show that these physical quantities differ strongly from the ones in standard percolation.
This explains the principal novelty of critical phenomena associated with this continuous transition and its surprising features including the small values of exponent $\beta$. Note that although the actual critical exponent of the order parameter is $m\beta$, Fig.~\ref{f5} shows that its value is anomalously small for all $m>1$. 
Note that this smallness is very unusual but 
%%actually 
not unprecedented. 
%%Interestingly, s
Small exponents of the order parameter were also observed in other 
non-equilibrium systems, 
%%transitions associated with 
in specific contact processes \cite{Odor:o04}. 
%%We completely described all scaling properties of this transition within the framework of the theory of continuous phase transitions.  
%%So there is no need to introduce new notions, like ``weakly discontinuous transition''.  
%%\cite{Nagler:nlt11}. 

The models which we considered were based on local optimization algorithms, in which each new connection requires a finite amount of information. Our work does not exclude possibility of discontinuity in more sophisticated models, where global optimization is implemented \cite{Araujo:ah10a,Schrenk:sfdadh12,Cho:chhk13,Chen:cncjszd13}. 
%%The linking rules of these models require global knowledge of a system or global control. 
To create new links in these models, one must know their global structure (i.e., all clusters) or globally control them.

%%knowledge of a percolation cluster or global control of the system. 

%% Refs.~\cite{Chen:czd12,Squires:ssaaog13}.  
 
%%Standard phase transition theory demonstrates how of continuous transitions may occur only in infinite systems~\cite{xxx}.
%%Concretely, the development of critical singularities requires the divergence of the correlation length, and in finite systems the correlation length can only be as large as the system's linear size.
%%In finite systems time $t$ (or the density of occupied nodes or edges) is a discreet quantity, as such, there are jumps of the order parameter for every percolation model, including ordinary percolation. If the jumps vanish in the infinite system size limit then it is a continuous percolation transition. Thus, there is no 
%%qualitative 
%%difference at all between continuous and weakly discontinuous transitions.***************

In this paper we focused on scaling properties of the transition but not on the critical point value $t_c$, which is of secondary interest for continuous phase transition and is determined by initial conditions (see Appendix~\ref{si10}). In our work \cite{daCosta:ddgm14} we showed how to obtain $t_c$ and $f(0)$ with high precision 
%%for several values of $m$ in our work \cite{xxx} for evolution started from isolated nodes. This method is applicable to 
for any initial conditions, and in Appendix~\ref{si11} we show how to get simple estimates. 

%% d upper  close to 2  -relation to simulation in 2d
%%By considering a set of models that extends over the entire range of explosive percolation problems we develop a generalized scaling theory of these systems. 
%%In particular, this theory concerns the class of explosive percolation transitions produced/driven by purely local optimization algorithms. 
%%It means that each new interconnection uses only a finite amount of information. 
%%It should be noticed that other more exotic models of explosive percolation employ global optimization rules, to those our theory may not apply. 
%%For example the models considered in \cite{xxx,xxx} feature bounded size rules which use an external variable to perform global control; and the algorithms in \cite{xxx,xxx} require the knowledge largest cluster's size present in the system at all times (to find the largest cluster, one has to know about all of them). This work does not exclude the possibility of discontinuities to occur in globally controlled systems.?

In summary, by developing the scaling theory for a wide class of models, we explained the continuous nature of the explosive percolation transition and its unusual properties. 
Our analysis can be extended to other systems and competition driven processes of this kind. 
We suggest that our work will provide a conceptual and methodological basis for new generalizations of percolation.

%%%% 
%%%%%%%%%%%%%%%%%%%%%%%%%%%%%%%%%%%%%%%%%%%%%%%%%%%%%%%%%%%%%%%%%%%%%%%
%%%%%%%%%%%%%%%%%%%%%%%%%%%%%%%%%%%%%%%%%%%%%%%%%%%%%%%%%%%%%%%%%%%%%%%%
%%\begin{figure}
%%%%%%[tbhd]
%%\begin{center}
%%%%\scalebox{1.00}{\includegraphics[angle=0]{log_tau-1_vs_m13.eps}}
%%\includegraphics[width=0.47\textwidth]{log_tau-1_vs_m13.eps}
%%\end{center}
%%\caption{ 
%%XXXXXXXXX
%%} 
%%\label{f4}
%%\end{figure}
%%%%%%%%%%%%%%%%%%%%%%%%%%%%%%%%%%%%%%%%%%%%%%%%%%%%%%%%%%%%%%%%%%%%%%%%
%%%%%%%%%%%%%%%%%%%%%%%%%%%%%%%%%%%%%%%%%%%%%%%%%%%%%%%%%%%%%%%%%%%%%%%%
%%%%
%%

%%\begin{acknowledgments} 
%%
%%This work was partially supported by the following projects PTDC: FIS/71551/2006, FIS/108476/2008, and SAU-NEU/103904/2008, and also by SOCIALNETS EU project. 
%%%%The authors thank M.~Barroso for his help with 
%%%%the 
%%%%administration of 
%%%%the
%%%%computing facilities used for numerical simulations. 
%%%%where the numerical calculations were done.
%%
%%\end{acknowledgments}

%%\begin{appendix}
%%
%%%%\section{The role of multiple connections}\label{appendix} 
%%%%\section{Pearson's coefficient of a growing network without multiple connections}\label{appendix}
%%\section{Exclusion of multiple connections}\label{appendix} 
%%%%connections} 

%%%%
%%\begin{equation}
%%...
%%\label{a10}
%%\end{equation} 
%%%%
 
%%\end{appendix}

\section*{Acknowledgements}

This work was partially supported by the FCT project PTDC/MAT/114515/2009 and
the FET proactive IP project MULTIPLEX number 317532. 
%%following projects PTDC: FIS/71551/2006, FIS/108476/2008, and SAU-NEU/103904/2008, and also by SOCIALNETS EU project. ...... ....... ........ ....... ......... ....... ........ ....... ....... .. ....... ........ ....... .......
%%The authors thank M.~Barroso for his help with 
%%the 
%%administration of 
%%the
%%computing facilities used for numerical simulations. 
%%where the numerical calculations were done.

%%%%%%%%%%%%%%%%%%%%%%%%%%%%%%%%%%%%%%%%%%%%%%%%%%%%%%%%%%%%
%%%%%%%%%%%%%%%%%%%%%%%%%%%%%%%%%%%%%%%%%%%%%%%%%%%%%%%%%%%%
%%%%%%%%%%%%%%%%%%%%%%%%%%%%%%%%%%%%%%%%%%%%%%%%%%%%%%%%%%%%

\section*{Author contributions}

All authors conceived and designed the research, worked out the theory, carried out the numerical calculations, analysed the results, and wrote the manuscript. 

%%%%%%%%%%%%%%%%%%%%%%%%%%%%%%%%%%%%%%%%%%%%%%%%%%%%%%%%%%%%
%%%%%%%%%%%%%%%%%%%%%%%%%%%%%%%%%%%%%%%%%%%%%%%%%%%%%%%%%%%%
%%%%%%%%%%%%%%%%%%%%%%%%%%%%%%%%%%%%%%%%%%%%%%%%%%%%%%%%%%%%

\section*{Additional information} 

The authors declare no competing financial interests. 
%%Supplementary information accompanies this paper on www....................... Reprints and permissions information is available online at http://.............. 
Correspondence and requests for materials should be addressed to S.N.D.

%%%%%%%%%%%%%%%%%%%%%%%%%%%%%%%%%%%%%%%%%%%%%%%%%%%%%%%%%%%%
%%%%%%%%%%%%%%%%%%%%%%%%%%%%%%%%%%%%%%%%%%%%%%%%%%%%%%%%%%%%
%%%%%%%%%%%%%%%%%%%%%%%%%%%%%%%%%%%%%%%%%%%%%%%%%%%%%%%%%%%%
%%%%%%%%%%%%%%%%%%%%%%%%%%%%%%%%%%%%%%%%%%%%%%%%%%%%%%%%%%%%
%%%%%%%%%%%%%%%%%%%%%%%%%%%%%%%%%%%%%%%%%%%%%%%%%%%%%%%%%%%%
%%%%%%%%%%%%%%%%%%%%%%%%%%%%%%%%%%%%%%%%%%%%%%%%%%%%%%%%%%%%

%%\newpage

%%\setcounter{equation}{0}
%%\setcounter{figure}{0}
%%\setcounter{section}{0}
%%\setcounter{table}{0}

%----------number equations as S1, S2... (since this is the Supplementary Information)
%%\renewcommand{\theequation}{S\arabic{equation}}
%%\renewcommand{\thefigure}{S\arabic{figure}}
%%\renewcommand{\thesection}{S\,\Roman{section}}
%%\renewcommand{\thetable}{S\,\Roman{table}}

%%\clearpage

\appendix

\bigskip

%%%%%%%%%%%%%%%%%%%%%%%%%%%%%%%%%%%%%%%%%%%%%%
%%%%%%%%%%%%%%%%%%%%%%%%%%%%%%%%%%%%%%%%%%%%%%
%%%%%%%%%%%%%%%%%%%%%%%%%%%%%%%%%%%%%%%%%%%%%%

\section{Contents of the Appendices}
\label{si0}

%%\section*{Supplementary\vspace{-6pt} Information}

%%\begin{center}
%%{\bf \vspace{10pt}Contents} 
%%\end{center}

\begin{itemize}

%\item[\ref{si0}.\ ]  
%Forbidding intra-cluster links

\item[\ref{si1}.\ ]  
Equations describing evolution of $S$, and $\langle s\rangle_P$  and $\langle s\rangle_Q$ 

\item[\hfill\ref{si2}.\ ]  
The nature of the order parameter and susceptibility

\item[\ref{si3}.\ ] 
Hyperscaling relations

\item[\ref{si4}.\ ]  
Derivation of equations for scaling functions

\item[\ref{si5}.\ ]  
Singularities of scaling functions at zero

\item[\ref{si3new}.\ ]
%%Scaling relation for the giant cluster size
Relation between the critical exponents $\beta$, $\tau$, and $\sigma$

\item[\ref{si6}.\ ]  
Asymptotics of scaling functions

\item[\ref{si7}.\ ]  
Solving equation for scaling functions numerically

\item[\ref{si8}.\ ]  
Critical exponents for $m$ from $1$ to $20$

\item[\ref{si9}.\ ]  
Above the transition

\item[\ref{si10}.\ ]  
Power-law initial conditions

\item[\ref{si11}.\ ]  
Simple estimates for the percolation \vspace{5pt} threshold

\end{itemize}

The following Appendices contain details of the calculations and arguments outlined or mentioned in the main text; a comprehensive and rigorous analysis of the observables for these problems, including the order parameter and susceptibility; discussion of generalizations to other models of this kind; and the table of precise numerical values for the critical exponents $\tau$ and $\beta$. 
In the following sections, when it is convenient, we reproduce some of the equations and formulas from the main text.

\section{Equations describing evolution of $S$, and $\langle s\rangle_P$  and $\langle s\rangle_Q$} 
\label{si1}

%%Equation~(\ref{e200}) leads to the following equations for the moments of the distributions and the size of the percolation cluster: 
The first of the Eqs.~(\ref{e300}) and~(\ref{e400}) in the main text, which we are reproducing here: 
\begin{eqnarray}
\frac{\partial S}{\partial t} 
&=  &
2 S^m \langle s \rangle_{\scriptscriptstyle \!Q} 
,
%%.
\label{se100}
%%\end{equation}
\\[5pt]
%%
%%and
%%
%%\begin{equation}
\frac{\partial \langle s \rangle_{\scriptscriptstyle \!P}}{\partial t} 
&= &
2 \langle s \rangle^2_{\scriptscriptstyle \!Q} - 
2 S^m \langle s^2 \rangle_{\scriptscriptstyle \!Q} 
, 
%%,
\label{se200}
\end{eqnarray}
%%
%%where $\langle s^n  \rangle_{\scriptscriptstyle \!P}=\sum_s s^nP(s)$ and $\langle s^n \rangle_{\scriptscriptstyle \!Q}=\sum_s s^nQ(s)$.
%%Equation~(3) 
demonstrates the principal difference of ``explosive'' percolation from ordinary one. Let us seed a giant component of relative size $h\ll 1$ in the normal phase at some moment $t<t_c$ 
%%, that is in the normal phase, 
and consider its 
%%immediate 
evolution. Equation~(\ref{se100}) shows that the growth rate of this component is proportional to $h^m$, i.e., it is severely suppressed in the entire normal phase if $m>1$. 
%%Only for ordinary percolation, the corresponding growth rate is proportional to the size $h$ of the seed component. 
This suppression results in the delayed transition compared to $m=1$. 
%%Equations~(\ref{e300}) and (\ref{e400}) will allow us to obtain conveniently relations between critical exponents. 

%%%%%%%%%%%%%%%%%%%%%%%%%%%%%%%%%%%%%%%%%%%%%%
%%%%%%%%%%%%%%%%%%%%%%%%%%%%%%%%%%%%%%%%%%%%%%
%%%%%%%%%%%%%%%%%%%%%%%%%%%%%%%%%%%%%%%%%%%%%%

\section{The nature of the order parameter and susceptibility} 
\label{si2}

Here we rigorously introduce the relevant order parameter and susceptibility and generalize them to other explosive percolation models.

In order to define the susceptibility for the explosive percolation problem, we use the relation between the susceptibility and correlation function that follows from the equivalence of the percolation problem to the one-state Potts model.
%\bibitem{KF}  P.W. Kosteleyn and C.M. Fortuin, J. Phys. Soc. Jpn. Suppl.
%{\bf 26}, 11 (1969); C.M. Fortuin and P.W. Kosteleyn, Physica {\bf 57}, 536
%(1972).
We define the correlation function $C(i,j)$ between vertices $i$ and $j$ as follows. Vertices $i$ and $j$ are correlated if they are connected by at least one path. In this case, $C(i,j)= 1$, otherwise $C(i,j)=0$. Moreover, by definition, $C(i,i)=1$. The susceptibility $\chi$ equals
\begin{equation}
\chi=\frac{1}{N}\sum_{i,j=1}^{N}C(i,j).
\label{os1}
\end{equation}
First we find the susceptibility for the ordinary percolation. In a system consisting of finite clusters, each labeled by index $\alpha$  and having size $s_{\alpha}$, Eq.~(\ref{os1}) takes a form,
\begin{equation}
\chi=N\sum_{\alpha} \Bigl(\frac{s_\alpha}{N} \Bigr)^2.
\label{os2}
\end{equation}
This equation shows that $\chi$ is related to the probability
$(s_\alpha/N)^2$
that two randomly chosen vertices belongs to the same cluster $\alpha$, i.e., they are connected.
The probability
\begin{equation}
x_\alpha=s_\alpha/N
\label{o1}
\end{equation}
that a randomly chosen vertex belongs to cluster $\alpha$ plays the role of an observable. In the phase without a giant component, all clusters are finite. Therefore,  at any $\alpha$,  $x_\alpha \rightarrow 0$ in the thermodynamic limit $N\rightarrow \infty$. In the phase with a giant cluster of relative size $S$, the corresponding observable $x_{gc}=S$ is nonzero and plays the role of the order parameter.

We can pass in Eq.~(\ref{os2}) from summation over individual clusters to summation over cluster sizes $s$, which gives for ordinary percolation
%%Defining $N(s)$ as the number of finite clusters of size $s$, we obtain
\begin{equation}
\chi
= N\sum_{\alpha} \Bigl(\frac{s_\alpha}{N} \Bigr)^2
=
N\sum_{s}\frac{NP(s)}{s} \Bigl(\frac{s}{N} \Bigr)^2
=
\sum_{s} P(s) s
,
\label{os4}
\end{equation}
%\begin{eqnarray}
%\chi & = & \sum_{s} \frac{N(s)}{N}s^2
%,
%\label{os3}
%\\[5pt]
% & = & \sum_{s} p(s) s
%,
%\label{os4}
%\end{eqnarray}
where $P(s) = N(s)s/N$
is the probability that a randomly chosen vertex belongs to a cluster of size $s$, $N(s)$ is the number of clusters of size $s$. Equation (\ref{os4}) is actually the standard definition of the susceptibility in ordinary percolation
%%introduces the standard definition of the susceptibility in the ordinary percolation problem
as the average size of the cluster to which a randomly chosen vertex belongs.

Let us now consider the explosive percolation problem. The
%%selection procedure
rule formulated in
%%Sec.~\ref{model}
the main text selects the vertex that belongs to the smallest of the $m$ clusters.
%%cluster of
%%chosen among $m$ clusters.
The probability that this vertex
%%belongs to
is in a cluster
%%with index
$\alpha$ equals
%%%%%%%%%%%%%%%%%%%%%%%%%%%%
%\begin{equation}
%x_\alpha=m\frac{s_\alpha}{N} \Bigl[ \sum_{s_\beta > s_\alpha} \frac{s_\beta}{N} \Bigr]^{m-1}
%\label{o2}
%,
%\end{equation}
%%%%%%%%%%
\begin{eqnarray}
&&
x_\alpha=m\frac{s_\alpha}{N} \Bigl[ \sum_{s_\beta > s_\alpha} \frac{s_\beta}{N} \Bigr]^{m-1}
\nonumber
\\[5pt]
&&
+ \frac{m(m-1)}{2!}\frac{s_\alpha}{N} \Bigl[\frac{N(s_\alpha) s_\alpha}{N}\Bigr] \Bigl[ \sum_{s_\beta > s_\alpha} \frac{s_\beta}{N} \Bigr]^{m-2}
\nonumber
\\[5pt]
&&
+ \dots +\frac{s_\alpha}{N}\Bigl[ \frac{N(s_\alpha) s_\alpha}{N} \Bigr]^{m-1}
\nonumber
\\[5pt]
&&
=\frac{1}{N(s_\alpha)}\sum_{k=1}^{m} \binom{m}{k} \Bigl[\frac{N(s_\alpha) s_\alpha}{N}\Bigr]^k \Bigl[ \sum_{s_\beta > s_\alpha} \frac{s_\beta}{N} \Bigr]^{m-k},
\label{o2}
\end{eqnarray}
%%
%where the summation is over all clusters $\beta$ of size $s_\beta$ larger than $s_\alpha$.
The first term is the probability that
the smallest of $m$ clusters, 
%%$\alpha$ 
%%to which a selected vertex belongs 
$\alpha$,  
has size $ s_\alpha$ while the other $m-1$ clusters are larger. The second term is the probability that apart from cluster $\alpha$ there is one more cluster of the same size while the remaining $m-2$ clusters 
%%have larger size, and so on. 
are larger. The last term is the probability that all $m$ clusters have the same size $ s_\alpha$. 
%%Finally, the sum of the probabilities is represented by the summation over $k$.

In the 
%%state without a percolation cluster, 
normal phase, the observables $x_\alpha \rightarrow 0$ in the thermodynamic limit $N\rightarrow \infty$. In the phase with a percolation cluster of the relative size $S$, the observable corresponding to the percolation cluster,
%$x_{gc}=S^m$
\begin{equation}
x_{\text{pc}}=S^m.
\label{o3}
\end{equation}
is nonzero and plays the role of the order parameter of the explosive percolation transition.

The susceptibility in the explosive percolation problem is a simple generalization of Eq.~(\ref{os2}) in which we replace the probability $s_\alpha/N$ by the probability Eq.~(\ref{o2}),
\begin{equation}
\chi=N\sum_{\alpha} x_{\alpha}^2.
\label{os5}
\end{equation}
This equation relates $\chi$ to the probability that two vertices chosen by use of the explosive percolation selection rules belong to the same cluster $\alpha$, i.e., they are connected. 
Relation (\ref{os5}) is valid in the normal phase, i.e., at $t< t_c$. In the phase with a percolation cluster at $t > t_c$, in Eq.~(\ref{os5}) we must subtract the contribution of the giant cluster,
\begin{equation}
\chi=N \Bigl[\sum_{\alpha} x_{\alpha}^2 - S^{2m} \Bigr].
\label{os7}
\end{equation}
%\begin{equation}
%\chi=m^2 \sum_{s} p(s) s \Bigl[ \sum_{s' \geq s} p(s') \Bigr]^{2(m-1)} - S^{2m}.
%\label{os7}
%\end{equation}
%%This equation is reduced to equation~(\ref{os6}) in which we must only sum over finite clusters, excluding the giant cluster.
%%One writes equation~(\ref{os5}) in a form
The right-hand sides of Eqs.~(\ref{os5}) and~(\ref{os7}) can be replaced with sums over $s$ accounting only for finite clusters. So both below and above $t_c$ we have
\begin{eqnarray}
%\chi=m^2 \sum_{s} p(s) s \Bigl[ \sum_{s' \geq s} p(s') \Bigr]^{2(m-1)}.
\chi
&=&
 N \sum_{s} \frac{1}{N(s_\alpha)} \left[ \sum_{k=1}^{m} \binom{m}{k}  P(s)^k  \Bigl[ \sum_{u \geq s+1} P(u) \Bigr]^{m-k}\right]^2 
 \nonumber
 \\[5pt]
&=&
\sum_s \frac{sQ(s)^2}{P(s)}
,
\label{os6}
\end{eqnarray} 
which coincides with Eq.~(\ref{e1800}). Here we used expression (\ref{e100}) for the distribution $Q(s)$.  
Equations (\ref{o3}) and (\ref{os7}) generalize the order parameter and susceptibility to the case of explosive percolation ($m >1$). 
 At $m=1$, these equations correspond to the ordinary percolation. At the critical point $t=t_c$, the susceptibility diverges, $\chi\rightarrow\infty$, manifesting the explosive percolation transition.

For the original Achlioptas process (product rule) the observable $x_{\alpha}$ is related to the probability that, 
%%at least in one of the two samplings, two vertices  belong to the same cluster:
a new link connects two vertices already in the same cluster:
\begin{eqnarray}
&&
x_{\alpha}^2 
= 
2 \Bigl(\frac{s_{\alpha}}{N} \Bigr)^2 \!\!\!\!\sum_{\beta,\gamma:\,s_{\beta} \times s_{\gamma} > s^2_{\alpha}} \!\!\frac{s_{\beta}}{N} \frac{s_{\gamma}}{N} 
\nonumber
 \\[5pt]
&&
+ 
 \Bigl(\frac{s_{\alpha}}{N} \Bigr)^2 \!\!\!\!\sum_{\beta,\gamma:\,s_{\beta} \times s_{\gamma} = s^2_{\alpha}} \!\!\frac{s_{\beta}}{N} \frac{s_{\gamma}}{N} 
 \label{se1300}
 \end{eqnarray}
The order parameter is $S^2$ as well as in our model. The susceptibility is given by Eq.~(\ref{os7}):
\begin{eqnarray}
&&
\chi=N \sum_{\alpha}  \Bigl[
2 \Bigl(\frac{s_{\alpha}}{N} \Bigr)^2 \!\!\!\!\sum_{\beta,\gamma:\,s_{\beta} \times s_{\gamma} > s^2_{\alpha}} \!\!\frac{s_{\beta}}{N} \frac{s_{\gamma}}{N}  
\nonumber
 \\[5pt]
&&
+
 \Bigl(\frac{s_{\alpha}}{N} \Bigr)^2 \!\!\!\!\sum_{\beta,\gamma:\,s_{\beta} \times s_{\gamma} = s^2_{\alpha}} \!\!\frac{s_{\beta}}{N} \frac{s_{\gamma}}{N}
\Bigr] .
\label{se1400}
 \end{eqnarray}
The first sum is over all clusters $\alpha$, excluding the giant component. 
For a general selection rule minimizing $f(s,s')$, 
the summation over $\beta,\gamma{:}\,s_{\beta} {\times} s_{\gamma} {> }s^2_{\alpha}$ and over $\beta,\gamma{:}\,s_{\beta}{ \times} s_{\gamma} {=} s^2_{\alpha}$ is replaced with summation over $\beta,\gamma{:}\,f(s_\beta,s_\gamma){>} f(s_\alpha,s_\alpha)$ and over $\beta,\gamma{:}\,f(s_\beta,s_\gamma){=} f(s_\alpha,s_\alpha)$, respectively.

The square of the order parameter is the probability that a new link is inside of the percolation cluster. For this, all four randomly chosen nodes must be in the percolation cluster, which gives $S^4$. That is, the order parameter for the  Achlioptas process is $S^2$ both for the product and sum rules (as well as for any other rule involving four nodes).

For the rule in which two optimal clusters from three are interlinked, we have the order parameter $S^{3/2}$. If this rule imposes selection of the pair with the smallest $f(s_\alpha,s_\beta)$, we have for the susceptibility:
%%
%\begin{equation}
%\chi=N \sum_{\alpha} \Bigl[ 3 \Bigl(\frac{s_{\alpha}}{N} \Bigr)^2
%\sum_{f(s_\alpha,s_\beta)\geq f(s_\alpha,s_\alpha)} \frac{s_{\beta}}{N} \Bigl]
%.
%\end{equation}
%%
%\begin{eqnarray}
%&&
%x_{\alpha}^2=3 \Bigl(\frac{s_{\alpha}}{N} \Bigr)^2
%\sum_{f(s_\alpha,s_\beta) > f(s_\alpha,s_\alpha)} \frac{s_{\beta}}{N}
%\nonumber
%\\[5pt]
%&&
%+ \Bigl(\frac{s_{\alpha}}{N} \Bigr)^3.
%\end{eqnarray}
\begin{equation}
\chi=
%%N \sum_{\alpha} x_{\alpha}^2=
N \sum_{\alpha} \Bigl[ 3 \Bigl(\frac{s_{\alpha}}{N} \Bigr)^2
\!\!\!\sum_{\beta:\,f(s_\alpha,s_\beta) > f(s_\alpha,s_\alpha)} \frac{s_{\beta}}{N}
+ \Bigl(\frac{s_{\alpha}}{N} \Bigr)^3 \Bigr]
.
\end{equation}

%%%%%%%%%%%%%%%%%%%%%%%%%%%%%%%%%%%%%%%%%%%%%%
%%%%%%%%%%%%%%%%%%%%%%%%%%%%%%%%%%%%%%%%%%%%%%
%%%%%%%%%%%%%%%%%%%%%%%%%%%%%%%%%%%%%%%%%%%%%%

\section{Hyperscaling relations} 
\label{si3}

Let us present hyperscaling relations for ordinary percolation (relations between scaling exponents including spatial dimensions) below the upper critical dimension $d_u$: 
\begin{eqnarray}
&& 
1/d_f =\sigma \nu
,
\label{se1400}
\\[5pt]
&&
d_f = d -\beta/\nu
,
\label{se1500}
\\[5pt]
&&
d-2+\eta = 2\beta/\nu
, 
\label{se1600}
\end{eqnarray}
where $d$ is the number of spatial dimensions and $d_f$ is the fractal dimension. 
The corresponding relations above 
%%To represent these Above 
$d_u$ are obtained by substituting $d_u$ for $d$, $1/2$ for  $\nu$, and $0$ for $\eta$. 
Here $1/2$ and $0$ are the mean-field theory values of the critical  exponents $\nu$ and $\eta$, respectively.
%% , change in these relations $d\to d_u$, $\nu \to 1/2$, and $\eta \to 0$. 

Let us recall how these relations were derived~\cite{Stauffer:sa-book94}. 

(i) Relation (\ref{se1400}). 

%%The largest finite cluster (the cutoff of the cluster size distribution $s_{\text{cut}}$ is of the order of 
According to the scaling form of the distribution  $P(s,t)$, see Eq.~(\ref{e500}), 
%%cluster sizes $s$ scale as $\delta^{-1/\sigma}$
the critical features are determined by cluster sizes  
%%Main contributions come from the cluster size 
$s \sim \delta^{-1/\sigma}$. In the critical region, the clusters are fractals, 
\begin{equation}
\delta^{-1/\sigma} \sim s \sim \xi^{d_f} \sim \delta^{-d_f\nu}
,
\label{se1700}
\end{equation}
where $\xi$ is the correlation length, $\xi \sim \delta^{-\nu}$, so we have relation (\ref{se1400}). One can see that this derivation is actually relevant for our explosive problem. 

(ii) Relation (\ref{se1500}). 

Consider a hyper-cube of $L^d$ nodes and estimate the number of nodes $M(L)$ of the percolation cluster 
falling inside this hyper-cube above $t_c$. 
%%Let $M(L)$ be the number of connected nodes in a hyper-cube of $L^d$. 
It is easy to  see that for $L \ll \xi$, this number is $M \sim L^{d_f}$, while for $L \gg \xi$ it is $M \sim S L^d$. 
So for $L\sim \xi$, 
\begin{equation}
\xi^{d_f} \sim \delta^\beta \xi^d
,
\label{se1800}
\end{equation}
which gives relation (\ref{se1500}). One can see that this derivation is also relevant for our explosive problem.  

(iii) Relation (\ref{se1600}). 

In general, the spin--spin correlation function near a continuous phase transition decays as $r^{-(d-2-\eta)}$ until the spin separation $r$ approaches the correlation radius $\xi$. 
%%where there is an exponential decay, 
So we can estimate 
\begin{equation}
\xi^{-(d-2-\eta)} \sim \phi^2 \sim \delta^{2\beta}
, 
\label{se1900}
\end{equation}
which gives relation (\ref{se1600}), if the order parameter $\phi = S \sim \delta^\beta$. 
For our model of explosive percolation, the order parameter $\phi = S^m $, i.e., $\phi \sim \delta^{\beta^*} \sim \delta^{m\beta}$, so for explosive percolation it should be 
\begin{equation}
d-2+\eta = 2\beta^*/\nu
.
\label{se2000}
\end{equation}
After substitution of the mean-field theory values $1/2$ for $\nu$, $0$ for $\eta$, and $d_u$ for $d$ we arrive at 
\begin{equation}
d_u-2 = 4\beta^* = 4m\beta
.
\label{se2100}
\end{equation}

%%On the other hand, if $m=2$, equation(\ref{e4020}) gives again  $d_u/2=1+4\beta$.

In addition we have 
\begin{eqnarray}
&& 
1/d_f =\sigma/2 = \frac{1}{2[1 + (2m-1)\beta]}
,
\label{se2200}
\\[5pt]
&&
d_f = d_u -2\beta
.
\label{se2300}
\end{eqnarray}
We emphasize that only two of the last three relations are independent. If, for example, we express $d_f$ and $d_u$ in terms of $\beta$ by using Eqs.~(\ref{se2100}) and~(\ref{se2200}) and then substitute the result into Eq.~(\ref{se2300}), we will arrive at the identity. 

 The upper critical dimension $d_u$ also describes the finite size effect for a continuous phase transition in systems above $d_u$, namely,  
%%The knowledge of $d_u$ allows us to describe 
%%%%the effect of finite size on the phase transition. 
%%%%Above the upper critical dimension, 
%%the finite size effect for a continuous phase transition. 
%%%%is described by the relation: 
%%Above the upper critical dimension, we have 
%%%%the finite size effect is
%%%%The knowledge of $d_u$ allows us to upper critical dimension allows us to describe the finite size effect, which is, a, described by the relation:  
%%
%%
%%derivation of relations between exponents 
%%
%%Critical exponents for explosive percolation are expressed in terms of exponent $\beta$: $\tau=1+\beta/(1+3\beta)$, $\sigma=1/(1+3\beta)$, $\gamma_{\scriptscriptstyle P}=(3-\tau)/\sigma=1+2\beta$, $\gamma_{\scriptscriptstyle Q}=(5-2\tau)/\sigma=1+\beta$, $d_f=2(1+3\beta)$, $d_u=2(1+4\beta)$. 
%%The size effect is described by the relation: 
%%
\begin{equation}
t_c(\infty)- t_c(N) \propto N^{-2/d_u}
.
%%, 
\label{e2400}
\end{equation}
%%
%%where $2/d_u=0....$. 
%%see Ref.~\cite{Privman:p90}. 
Here $t_c(\infty)$ is the critical point value in the infinite system (in which the transition is well defined) and $ t_c(N)$ is, in particular, the position of the maximum of the susceptibility for the system of $N$ nodes.

%%%%%%%%%%%%%%%%%%%%%%%%%%%%%%%%%%%%%%%%%%%%
%%%%%%%%%%%%%%%%%%%%%%%%%%%%%%%%%%%%%%%%%%%%
%%%%%%%%%%%%%%%%%%%%%%%%%%%%%%%%%%%%%%%%%%%%

\section{Derivation of equations for scaling functions} 
\label{si4} 

In this section we show in detail how to derive equations for scaling functions from the evolution equations. 
We suggest that the ideas implemented in this derivation will be useful for numerous generalizations of percolation. 

%%We are interested not ....
In our 
%%previous 
work of Ref.~\cite{daCosta:ddgm10}, we have shown that if it is known that the distribution $P(s)$ at the critical point is, asymptotically, power-law with some given critical exponent and amplitude, $P(s,t_c) \cong A s^{-\tau+1}$, as it should be for a continuous phase transition, then from Eq.~(\ref{e200}), immediately follows the power law $S \cong B \delta^\beta$, where 
$\beta = (\tau-2)/[1 - (2m-1)(\tau-2)]$ 
as in Eq.~(\ref{e1400}) and the coefficient $B$ is expressed in terms of $A$ and $\tau$. 
(Here the critical amplitude $A$ is determined by the initial form of the distribution $P(s,t=0)$.)
Furthermore, this assumption allows us to find the scaling functions $f(x)$ and $g(x)$ on the upper side of the phase transition, i.e. at $t>t_c$. The form of these functions turns out to be close to exponential, similarly to ordinary percolation above an upper critical dimension. The derivation detailed in Appendix~\ref{si9} exploits the convenient simplification of the equations above the critical point, where $S$ differs from zero. In this region, at large $s$, Eq.~(\ref{e100}) is reduced asymptotically to $Q(s) \cong mS^{m-1}P(s)$, which makes the resulting evolution equation for $P(s,t)$ to be similar to that for ordinary percolation and so easily solvable with the initial condition $P(s,t_c) \cong A s^{-\tau+1}$. 
Therefore our present more difficult task is to find the distribution at the critical point, which we just used in that derivation, its critical exponent (if this distribution will appear to be power-law), and the scaling functions on the normal phase side of the phase transition, i.e. at $t<t_c$. So, simultaneously we verify that the transition is continuous.  

First we 
%%need to 
derive equation for scaling functions approaching the critical point from the normal-phase side by using Eqs.~(\ref{e200}) and (\ref{e800}). 
The direct substitution of the scaling forms of the distributions $P(s,\delta)$ and $Q(s,\delta)$ into the evolution Eq.~(\ref{e200}) is impossible, since these forms are valid at large $s$, while the contribution from the region of small $s$ to the sum in Eq.~(\ref{e200}) is nonzero. 
%%(actually divergent). 
%%This does not allow us to directly substitute the sum with an integral
%%The difficulty is that, into the evolution equation, we must substitute 
%%Into the evolution equation (\ref{e100}) and the evolution equation (\ref{e200}) we must substitute 
%%the scaling forms of the distributions $P(s,\delta)$ and $Q(s,\delta)$, which are valid at large $s$ 
%%and small deviations $\delta$ from the critical point. 
%%We must substitute the scaling forms The scaling forms of the distributions 
Let us rewrite Eq.~(\ref{e200}) to eliminate this contribution from the sum and so to remove the non-scaling, low $s$ parts of the distribution from consideration. We substitute $Q(u) = Q(s) + [Q(u)-Q(s)]$ into the evolution equation, which leads to the following equation:   
\begin{eqnarray}
&& 
\!\!\!\!\!\!\!\!\!
\frac{\partial P(s)}{\partial t} = -s(s-1) Q^2(s) + 2sQ(s)[1 - \sum_{u=s}^{\infty} Q(u)] 
%%\sum_{u=1}^{s-1}Q(u) 
\nonumber
\\[5pt]
&& 
\!\!\!\!\!\!\!\!\!
+ s \sum_{u=1}^{s-1} [Q(u) - Q(s)][Q(s-u) - Q(s)] - 2sQ(s)  
, 
\label{e2500}
\end{eqnarray}
in which we 
%%now 
can safely substitute integrals for the sums. The resulting equation is 
\begin{eqnarray}
&& 
\frac{\partial P(s)}{\partial t} \cong 
-s^2 Q^2(s) - 2s Q(s) \int_s^\infty \!\! du\, Q(u) 
\nonumber
\\[5pt]
&& 
+ 
s\int_0^s du\, [Q(u) - Q(s)][Q(s-u) - Q(s)]
. 
\label{e2600}
\end{eqnarray}
The scaling form of the distribution $P(s,t)$ for large $s$ in the critical region is  
\begin{equation}
P(s,t) = s^{1-\tau}f(s\delta^{1/\sigma}) = \delta^{(\tau-1)/\sigma}\tilde{f}(s\delta^{1/\sigma})
, 
%%,
\label{se2610}
\end{equation}
where $\delta=|t-t_c| \ll 1$, and $f(x)$ and $\tilde{f}(x)$ are scaling functions, $f(x) = x^{\tau-1}\tilde{f}(x)$,  $\tau$ and $\sigma$ are critical exponents. 
These two functions, $f(x)$ and $\tilde{f}(x)$, provide two equivalent representations of scaling. 
In the following, $\tilde{f}(x)$ turned out to be more convenient for us. 
%%For the size distribution of clusters at the critical point we have $n(s) \sim s^{-\tau}$. 
%%
On the other hand, the scaling form of the $Q(s,t)$ distribution is 
%%
%%$$
%%Q(s,t) = s^{(m-1)-m\tilde{\tau}}g(s\delta^{1/\sigma}) = \delta^{[m\tilde{\tau}-(m-1)]/\sigma}\tilde{g}(s\delta^{1/\sigma})
%%$$ 
%%
\begin{equation}
Q(s,t) = s^{(2m-1)-m\tau}g(s\delta^{1/\sigma}) = \delta^{[m\tau-(2m-1)]/\sigma}\tilde{g}(s\delta^{1/\sigma})
,   
\label{se2700}
\end{equation}
where $g(x) = x^{m\tau-(2m-1)}\tilde{g}(x)$. 
Substituting these scaling forms of the distributions into Eqs.~(\ref{e2600}) and (\ref{e800}), and equating the powers of $\delta$ in all the terms, we arrive at the following equation for the scaling functions: 
\begin{eqnarray}
&&  
- \frac{\tau-1}{\sigma}\tilde{f}(x) - \frac{1}{\sigma}x\tilde{f}'(x) 
\nonumber
\\[5pt]
&&
= - x^2\tilde{g}^2(x) - 2x\tilde{g}(x) \int_x^\infty dy\, \tilde{g}(y) 
\nonumber
\\[5pt]
&&
+ x \int_0^x dy\, [\tilde{g}(y) - \tilde{g}(x)][\tilde{g}(x-y) - \tilde{g}(x)]
\label{se2800}
\\[5pt]
&&
%%\int_x^\infty dy\, y^{ggg}\tilde{g}(y) = \Biggl[\int_x^\infty dy\, y^{ggg}\tilde{f}(y)\Biggr]^m
\tilde{g}(x) = m\Biggl[\int_x^\infty dy\,\tilde{f}(y)\Biggr]^{m-1} \tilde{f}(x)
, 
\label{se2900}
\end{eqnarray}
where 
\begin{equation}
\sigma = 1 - (2m-1)(\tau-2) 
%%= 1/[1 + (2m-1)\beta]
. 
\label{se2950}
\end{equation}
The last relation for the critical exponents follows from the condition that all factors containing powers of $\delta$ must cancel each other. 
Equation~(\ref{se2800}), with substituted $\tilde{g}(x)$ from Eq.~(\ref{se2900}) can be treated as a nonlinear integral differential eigenfunction equation for the scaling function $\tilde{f}(x)$, in which a critical exponent, say $\tau$, plays the role of the eigenvalue. 
Note that Eqs.~(\ref{se2800}) and (\ref{se2900})
%%Note that the right-hand side of this equation 
inconveniently contain integrals with integration over different intervals, $(x,\infty)$ and $(0,x)$. 
%%For solving the equation, we 
To avoid this inconvenience, we must exclude the integrals over the interval $(x,\infty)$. 
%%For that, we take the derivative of both sides of equation~(\ref{e2800}) and substitute. 
For that, in both Eqs.~(\ref{se2800}) and (\ref{se2900}) we 
%%separate the integrals over $(x,\infty)$ 
move the integrals over $(x,\infty)$ to the left-hand sides of the equations and move everything else to the right-hand sides, 
%%, express them in terms of the rest parts of the equations and 
and then take the derivatives of the both sides. 
%%of each of resulting equations. 
The derivation removes the integrals $\int_x^\infty$, but, unfortunately, produces new divergencies within the remaining integrals $\int_0^x$. To avoid these divergencies, it is sufficient first to pass from  
%%In addition, it is necessary  
%%%%have 
%%to transform 
the integral over the interval $(0,x)$ to integration over $(0,x/2)$ in Eq.~(\ref{se2800}), namely 
\begin{eqnarray}
&&
\int_0^x dy\, [\tilde{g}(y) - \tilde{g}(x)][\tilde{g}(x-y) - \tilde{g}(x)] 
\nonumber
\\[5pt]
&&
= 2 \int_0^{x/2} dy\, [\tilde{g}(y) - \tilde{g}(x)][\tilde{g}(x-y) - \tilde{g}(x)]
.
\label{se3200}
\end{eqnarray}
%%
%%(to avoid singularities). 
The resulting system of two equations contains $\tilde{f}''(x)$, $\tilde{f}'(x)$, $\tilde{f}(x)$, $\tilde{g}'(x)$, and $\tilde{g}(x)$. Introducing $\tilde{u}(x) = \tilde{f}'(x)$, we obtain the system of three first order equations for $\tilde{f}(x)$, $\tilde{g}(x)$, and $\tilde{u}(x)$:   
\begin{eqnarray}
&&
\tilde{f}''(x) = \tilde{u}'(x) = 
\frac{\tau-1}{x}\Bigl[\frac{\tilde{f}(x)}{x} - \tilde{f}'(x)\Bigr] 
\nonumber
\\[5pt]
&&
+ \frac{\tilde{g}'(x)}{\tilde{g}(x)}\Bigl[\frac{(\tau-1) \tilde{f}(x)}{x} + \tilde{f}'(x)\Bigr] - 
\sigma \tilde{g}^2(x/2) 
\nonumber
\\[5pt]
&&
+ 
\frac{2\sigma}{\tilde{g}(x)} \int_0^{x/2} \!\!\!\!\!dy\, \tilde{g}(y)[\tilde{g}'(x)\tilde{g}(x-y) - \tilde{g}(x)\tilde{g}'(x-y)]
\nonumber
\\[5pt]
&&
\tilde{g}'(x) =\frac{\tilde{f}'(x)\tilde{g}(x)}{\tilde{f}(x)} - m(m-1)\tilde{f}^2(x)\Bigl [\frac{\tilde{g}(x)}{m\tilde{f}(x)}\Bigr]^{(m-2)/(m-1)}
\nonumber
\\[5pt]
&&
\tilde{f}'(x) = \tilde{u}(x)
, 
\label{se3300}
\end{eqnarray}
where the exponent $\sigma$ is related with $\tau$ according to (\ref{se2950}). 
%%$\sigma = 1 - (2m-1)(\tau-2) = 1/[1 + (2m-1)\beta]$. 

%%%%%%%%%%%%%%%%%%%%%%%%%%%%%%%%%%%%%%%%%%%%
%%%%%%%%%%%%%%%%%%%%%%%%%%%%%%%%%%%%%%%%%%%%
%%%%%%%%%%%%%%%%%%%%%%%%%%%%%%%%%%%%%%%%%%%%

\section{Singularities of scaling functions at zero} 
\label{si5} 

One can verify that at small $x$, the solution of this system has the following expansion: 
%%we find the following One can check This system 
%%
\begin{eqnarray}  
&& 
\!\!\!\!\!\!\!\!\!\!\!\!\!
f(x) = x^{\tau-1}\tilde{f}(x) = f(0) + a_1 x^\sigma + a_2 x^{2\sigma} + ...
, 
\nonumber
\\[5pt]
&&
\!\!\!\!\!\!\!\!\!\!\!\!\!
g(x) = x^{m\tau-(2m-1)}\tilde{g}(x) = g(0) + b_1 x^\sigma + b_2 x^{2\sigma}{+} ...
, 
\label{se3000}
\end{eqnarray}
where $f(0)$ and $g(0)$ are the critical amplitudes of the distributions, $P(s,t_c) \cong f(0)s^{1-\tau}$ and $Q(s,t_c) \cong g(0)s^{(2m-1)-m\tau}$, respectively. 
%%Let us discussw the way we 
One can easily find that $g(0)$ and all other coefficients in these series are expressed in terms of only $f(0)$ and $\tau$. For example, from the relation
\begin{equation}
\int_x^\infty \! dy \,y^{(2m-1)-m\tau} g(y) = \Biggl[\int_x^\infty \! dy \,y^{1-\tau} f(y)\Biggr]^m
, 
\label{se3500}
\end{equation}
we immediately obtain 
\begin{equation}
g(0) = \frac{m}{(\tau-2)^{m-1}}f^m(0)
. 
\label{se3600}
\end{equation}
Similarly, we obtain the next coefficients using relations~(\ref{se2800}) and~(\ref{se3500}), 
%%The next coefficients are
%%
\begin{eqnarray}
%%&&
%%\!\!\!\!\!\!\!\!\!\!\!\!\!
%%a_1=-\frac{2 g(0)^2 \pi  \csc[-m \pi(\tau-2)] \Gamma[1-m \left( \tau-2 \right)]}{\Gamma[1+m \left(\tau-2 \right)] \Gamma[1-2 m (\tau-2)]}
%%,
%%\nonumber
%%\\[5pt]
&&
\!\!\!\!\!\!\!\!\!\!\!\!\!
a_1=-\frac{ g(0)^2 \Gamma[-m \left( \tau-2 \right)]^2}{\Gamma[-2 m (\tau-2)]}
,
\nonumber
\\[5pt]
%%&&
%%\!\!\!\!\!\!\!\!\!\!\!\!\!
%%b_1=g(0) a_1 \left(\frac{1}{f(0)}+\frac{(m-1)\left(\frac{g(0)}{f(0) m}\right)^{1/(1-m)}}{2 m (\tau-2) - 1}\right)
%%, 
%%\nonumber
%%\\[5pt]
&&
\!\!\!\!\!\!\!\!\!\!\!\!\!
b_1=\frac{a_1 g(0)[1-(3 m-1) (\tau-2)] }{f(0) [1-2 m (\tau-2)]}
,
\nonumber
\\[5pt]
%%&&
%%\!\!\!\!\!\!\!\!\!\!\!\!\!
%%a_2=\frac{a_1 b_1}{2g(0)} +\frac{g(0) b_1 \pi \csc\left[-m\pi (\tau-2)\right]}{\Gamma[1+m(\tau-2)]}
%%\nonumber
%%\\
%%&&
%%\!\!\!\!\!\!\!\!\!\!\!\!\!
%%\times \left( \frac{4^{m(\tau-2)}\sqrt{\pi}}{\Gamma[1/2-m(\tau-2)]} -\frac{\Gamma[1-(3m-1)(\tau-2)]}{\Gamma[1-(4m-1)(\tau-2)} \right)
%%, 
%%\nonumber
%%\\[10pt]
&&
\!\!\!\!\!\!\!\!\!\!\!\!\!
a_2=\frac{a_1 b_1}{2g(0)} +g(0) b_1\Gamma[-m(\tau-2)]
\nonumber
\\
&&
\!\!\!\!\!\!\!\!\!\!\!\!\!
\times \left( \frac{4^{m(\tau-2)}\sqrt{\pi}}{\Gamma[1/2-m(\tau-2)]} -\frac{\Gamma[1-(3m-1)(\tau-2)]}{\Gamma[1-(4m-1)(\tau-2)} \right)
, 
\nonumber
\\[10pt]
&&
\!\!\!\!\!\!\!\!\!\!\!\!\!
b_2=\frac{g(0)[(5m-2)(\tau-2)-2]}{f(0)}
\nonumber
\\
&&
\!\!\!\!\!\!\!\!\!\!\!\!\!
\times \left(\frac{a_1^2(m-1)(\tau-2)}{2f(0)[1-2m(\tau-2)]^2}-\frac{a_2}{\tau-4m(\tau-2)}\right) 
, 
\label{se3700}
\end{eqnarray}
%%
%%%%%\begin{align}
%%%%%a_{g}&= {a_{f}}^m m (\tau-1)^{1-m}
%%%%%\\[10pt]
%%%%%b_{f}&=\frac{2 a_{f}^2 \pi  \csc[-m \pi\left( \tau-1 \right)] \Gamma[1-m \left( \tau-1 \right)]}{\Gamma[1+m \left(\tau-1 \right)] \Gamma[1-2 m \left( \tau -1\right)]}
%%%%%\\[10pt]
%%%%%b_{g}&=a_{g} b_{f} \left(\frac{1}{a_{f}}+\frac{\left( m-1 \right) \left(\frac{a_{g}}{a_{f} m}\right)^{\frac{1}{1-m}}}{2 m \left(\tau -1 \right)-1}\right)
%%%%%\end{align}
%%%%%\\[10pt]
%%%%%\\
%%%%%$\Gamma\equiv$ Gamma Funtion\\
%$\gamma\equiv$ Euler-Mascheroni Constant $=0.5772156649...$\\
%$\Psi\equiv$ Digamma Function or Zeroth Order Polygamma Function\\
%%$\csc\equiv$ Cosecant Function
and so on. 
We do not show here the next two pairs of coefficients $a_k$ and $b_k$ which we have also obtained using Mathematica since they are too cumbersome. 
%%ADD THE NEXT TERMS .... 
If we know $f(0)$ and $\tau$, these Taylor series 
%%, converging in the range below the maxima of the scaling functions, 
provide the solutions $f(x)$ and $g(x)$ only at sufficiently small $x$ (below the maxima of the scaling functions). 
%%In this way we construct the Taylor series of the solutions in $x$, which converges in the range below the maxima of the scaling functions. 

Note that the case of ordinary percolation, i.e. $m=1$, is special for series (\ref{se3000}) in the following sense. It turns out that for $m=1$, the odd coefficients in these series are zero. For example, one can easily check in Eq.~(\ref{se3700}) that in this case, $a_1=b_1=0$. This feature also follows from the form of the scaling function for ordinary percolation, $f(x) = e^{-2x}/\sqrt{2\pi}$.

%%%%%%%%%%%%%%%%%%%%%%%%%%%%%%%%%%%%%%%%%%%
%%%%%%%%%%%%%%%%%%%%%%%%%%%%%%%%%%%%%%%%%% New section

%%\section{Scaling relation for the giant cluster size}
\section{Relation between the critical exponents $\beta$, $\tau$, and $\sigma$}
\label{si3new}

Let us 
%%find scaling behavior 
analyze the critical singularity of the giant cluster size $S$. 
%%near the critical point, 
Let deviations from the critical point be small, $\delta=t -t_c \ll t_c$. Substituting the scaling form of the distribution $P(s,t)$, 
%%function 
Eq.~(5), into Eq.~(6) and replacing 
%%the 
summation 
%%to the 
with integration, we find
\begin{eqnarray}
&&
S \approx \int_{1}^{\infty} s^{1-\tau}[f(0)-f(s\delta^{1/\sigma})]ds
\nonumber
 \\[5pt]
&&
=\delta^{(\tau-2)/\sigma} \int_{\delta^{1/\sigma}}^{\infty} x^{1-\tau}[f(0)-f(x)]dx.
\label{si3n-1}
\end{eqnarray}
%%
%%The i
Integration by parts leads to
\begin{equation}
S \approx \frac{\delta^{(\tau-2)/\sigma}}{\tau-2} \Bigr\{\!\!- x^{2-\tau}[f(0){-}f(x)] \Big|_{0}^{\infty} 
- \int_{0}^{\infty}\!\!\! x^{2-\tau}\frac{f(x)}{dx} dx \Bigl\}.
\label{si3n-2}
\end{equation}
This equation shows that $S \sim \delta^{\beta}$ with the critical exponent 
$$
\beta=(\tau-2)/\sigma
$$ 
%(see Eq.~(\ref{e600})) 
if the scaling function $f(x)$ satisfies the following conditions. First, $x^{2-\tau}[f(0){-}f(x)] {\to} 0$ at $ x {\to} 0$. Second, the integral in Eq.~(\ref{si3n-2}) is finite. These assumptions impose conditions on the value of $\tau$ and the behavior of $f(x)$ at $x \ll 1$ and $x \gg 1$ \cite{Stauffer:s79}. In particular, using the lowest term of the series %%Eq.~
(\ref{se3000}), we find that the equality $2 < \tau < 2+\sigma$ must be satisfied at $m \geq 2$ in contrast to $2 < \tau < 3$ for ordinary percolation ($m=1$) \cite{Stauffer:s79}. The inequality $ \tau < 2+\sigma$ substituted into relation (\ref{se2950}) for the critical exponents $\sigma$ and $\tau$ results in the condition $2 < \tau < 2+1/(2m)$ for $m\geq2$. 
(As we mentioned above, the case of $m=1$ is special here, since in this case, the odd coefficients in the series (\ref{se3000}) are zero.) 
Our solution presented 
%%for $m \geq 2$
 in Appendix~\ref{si5} and in 
%%the 
Table \ref{st1} satisfies these conditions. 
%%It 
This evidences the self-consistency of the solution and the assumptions used when obtaining the scaling relations. 
%%Note that the 

%%%%%%%%%%%%%%%%%%%%%%%%%%%%%%%%%%%%%%%%%%%%
%%%%%%%%%%%%%%%%%%%%%%%%%%%%%%%%%%%%%%%%%%%%
%%%%%%%%%%%%%%%%%%%%%%%%%%%%%%%%%%%%%%%%%%%%

\section{Asymptotics of scaling functions} 
\label{si6} 

Let us find 
%that at large $x$
the asymptotic behavior of the scaling functions explicitly. Tending $x\to \infty$, and taking the leading terms on each side of Eqs.~(\ref{se2800}) and~(\ref{se2900}), 
one can easily check that they have the following rapidly decaying asymptotics:
\begin{align}
&\tilde{f}(x) \cong A x^{\lambda}\exp \left[{-Cx^{1+\ln m/\ln 2}}\right]
,
\nonumber 
\\[5pt]
&\tilde{g}(x)\cong \frac{mA^m x^{m\lambda - (m-1)\ln m/\ln 2}}{\left[C\left(1+\ln m/\ln 2\right)\right]^{m-1}}\exp \left[{-mCx^{1+\ln m/\ln 2}}\right]
,
\label{eq:f_g_m_asymp}
\end{align}
where 
$$
\lambda=\left(1+\frac{\ln m}{\ln 2}\right)\left(1+\frac{1}{4m-2}\right) - \frac{2m}{2m-1}
.
$$ 
This procedure also gives a relation between constants $A$ and $C$,
\begin{eqnarray}
&&
\hspace{-15pt}
A^{2m-1}=\left(\frac{\ln m}{\pi \ln 2}\right)^{1/2}
\nonumber
\\[5pt]
&&
\hspace{-15pt}
\times
\frac{(1+\ln m/\ln 2)^{2m-1/2} 2^{\lambda+1+\ln m/\ln 2} }{\sigma m^{3/2}} C^{2m-1/2},
%A^{2m-1}=\left(\frac{\ln m}{\pi \ln 2}\right)^{1/2}\frac{\left[C(1+\ln m/\ln 2)\right]^{2m-1/2} 2^{\lambda+1+\ln m/\ln 2} }{\sigma m^{3/2}}
%%\nonumber
\label{ee}
\end{eqnarray}
however, does not fix them. $A$ and $C$ are determined by $f(0)$, which is in its turn determined by  the initial distribution $P(s,t=0)$.

\section{Solving equations for scaling functions numerically} 
\label{si7} 

%%and so on. 
%%If we know $f(0)$ and $\tau$, 
The Taylor series (\ref{se3000}) 
%%, converging in the range below the maxima of the scaling functions, 
provide the solutions $f(x)$ and $g(x)$ at small $x$ in terms of yet unknown $f(0)$ and $\tau$. 
%%(below the maxima of the scaling functions). 
%%In this way we construct the Taylor series of the solutions in $x$, which converges in the range below the maxima of the scaling functions. 
The critical amplitude $f(0)$, as well as the detailed shapes of the scaling functions, are determined by the initial distribution of cluster sizes, 
%%at $t=0$, 
$P(s,t=0)$. In contrast to that, the critical exponent values do not depend on 
%%the 
initial conditions, if $P(s,t=0)$ decays sufficiently rapidly (see below). 
%%(Note however that, unlike the critical exponents, the scaling function shapes depend on the initial conditions.)   
So, 
%%while 
when searching for the solution of the system of Eqs.~(\ref{se3300}),  
%%Since the resulting value of the critical exponent $\tau$ is the same for a wide range of initial conditions (the distribution $P(s,t=0)$ must decay faster than ... finite susceptibility ), and we are interested only in $\tau$, 
we can set any convenient value of the critical amplitude $f(0)$. 
%%to find the solution of the equation. 
For different values of $f(0)$, the resulting value of the critical exponent $\tau$ should be the same, and the scaling functions, while differing from each other, should be qualitatively 
%%very 
similar. 
For a given critical amplitude $f(0)$, the system of first order differential Eqs.~(\ref{se3300}) can be 
%%easily 
directly solved numerically. This solution should give the exponent $\tau$ together with the scaling functions $f(x)$ and $g(x)$. 
%%for a given $f(0)$. 
%%To .....   condition
The unknown critical exponent $\tau$ is obtained from the condition that $f(x)$ and $g(x)$ decay rapidly to zero as $x$ approaches infinity. 

We use the following 
%%numerical 
procedure. 
For the sake of convenience, set the value of the critical amplitude $f(0)$ such that the maxima $f(x)$ and $g(x)$ are of the order of $1$ 
%%(it turnes out that 
(with this choice, the numerical solution 
%%of the system (\ref{e2900}) 
takes minimum time). First try some reasonable value of $\tau$. 
Insert this pair, $f(0)$ and $\tau$ into truncated series (\ref{se3000}) 
%%with coefficients (\ref{se3600}), (\ref{se3700}),  (\ref{se3xxx}), and so on 
and use them at some small $x_0$ as  initial conditions for the first order Eqs.~(\ref{se3300}). 
With these initial conditions, 
%%With this pair, $f(0)$ and $\tau$, using truncated series (\ref{se3000}) with coefficients (\ref{se3600}), (\ref{se3700}),  (\ref{se3xxx}), and so on as initial conditions at some 
%%sufficiently 
%%small $x_0$, 
find the numerical solution of the system 
%%of the first order equations 
(\ref{se3300}) up to sufficiently large $x$ 
%%within range of asymptotic .... 
at which the asymptotics of the solutions are already visible. 
%%asymptotic behaviors of the solutions are already clear. 
Since the value of $\tau$, which we used in this first attempt, 
%%of course 
surely deviates from the correct one, the obtained solutions will not show a proper 
%%rapid 
decay to zero. Instead, they may decay more slowly than exponentially or even become negative, oscillate, and so on. Then solve equation numerically with a different value of $\tau$, and repeat this procedure again and again, adjusting progressively the value of $\tau$ in such a way that the solutions $f(x)$ and $g(x)$ decay to zero more and more rapidly, staying positive. These calculations converge rapidly giving the final value of $\tau$ with any desired precision and the scaling functions $f(x)$ and $g(x)$, see Fig.~\ref{f4}.

%%%%%%%%%%%%%%%%%%%%%%%%%%%%%%%%%%%%%%%%%%%%
%%%%%%%%%%%%%%%%%%%%%%%%%%%%%%%%%%%%%%%%%%%%
%%%%%%%%%%%%%%%%%%%%%%%%%%%%%%%%%%%%%%%%%%%%

\section{Critical exponents for $m$ from $1$ to $20$} 
\label{si8} 

The list of values of the exponent $\tau$ for $m$ from $1$ to $20$ plotted in Fig.~\ref{f5} is presented in Table~\ref{st1}. These values were obtained in the way described in Appendix~\ref{si7}. Table~\ref{st1} also contains the values of exponent $\beta$, obtained from $\tau$ using the following relation: 
$$
\beta = \frac{\tau - 2}{1 - (2m-1)(\tau-2)}
.
$$

%%From phase diagram if initial condition - bare nodes: 
%%
%%$m=2$:
%%
%%$t_c =0.92320750930 (2)$,
%%
%%$\tau=2.04763044 (2)$,
%%
%%$f(0)=0.04619068 (2)$,
%%
%%${\cal P}(1,t_c)=0.04859280 (1)$.
%%
%%$m=3$:
%%
%%$t_c =0.98179531735 (5)$,
%%
%%$\tau=2.00991188 (2)$,
%%
%%$f(0)=0.00983139 (1)$,
%%
%%${\cal P}(1,t_c)=0.011721464802 (3)$.
%%
%%$m=4$:
%%
%%$t_c =0.99497356260 (5)$,
%%
%%$\tau=2.00243833 (1)$,
%%
%%$f(0)=0.00244 (2)$,
%%
%%${\cal P}(1,t_c)=0.003343067143 (3)$. 
%%
%%
%%\noindent 
%%$
%%m \ \ \ \ \ \ \ \ \ \ \tau 
%%\\
%%2 \ \ \ 1.04763044(2)
%%\\
%%3 \ \ \ 1.00991188(2)
%%\\
%%4 \ \ \ 1.0024383316(5)
%%\\
%%5 \ \ \ 1.000625199(1)
%%\\
%%6 \ \ \ 1.0001601191(4)
%%\\
%%7 \ \ \ 1.00004044602(2)
%%\\
%%8 \ \ \ 1.00001006831(3)
%%\\
%%9 \ \ \ 1.00000247685(2)
%%\\
%%10 \ \ 1.00000060412(6)
%%\\
%%11 \ \ 1.000000146392(2)
%%\\
%%12 \ \ 1.000000035313(2)
%%\\
%%13 \ \ 1.000000008489(2)
%%\\
%%14 \ \ 1.0000000020355(2)
%%\\
%%15 \ \ 1.0000000004870(1)
%%\\
%%16 \ \  1.00000000011634(4)
%%\\
%%17 \ \  1.00000000002776(2)
%%$

%%%%%%%%%%%%%%%%%%%%%%%%%%%%%%%%%%%%%%%%%%%%%%%%%
%%%%%%%%%%%%%%%%%%%%%%%%%%%%%%%%%%%%%%%%%%%%%%%%%

\begin{table}
\caption[Solution for $m=2,3,\dots,20$]{Critical exponents $\tau$ and $\beta$ for $m$ from $1$ to $20$.} 
\begin{center}
\begin{tabular}{l|l|l}
\hline
\\[-12pt]
%\noalign{\smallskip}
$m$     &  $\tau$   &  $\beta$  
\\[2pt] \hline\hline
\\[-11pt]
$1$   & 2.5   &      1
\\[1pt]
$2$   & 2.04763044(2)       & $5.557106(2) \times 10^{-2}$
\\[1pt]
$3$   & 2.00991188(1)       &   $1.042872(1) \times 10^{-2}$
\\[1pt]
$4$   & 2.002438330(5)       &   $2.480671(5) \times 10^{-3}$
\\[1pt]
$5$   & 2.000625199(1)       &   $6.28737(1)\times 10^{-4}$
\\[1pt]
$6$   & 2.0001601191(4)      &    $1.604016(4)\times 10^{-4}$
\\[1pt]
$7$   & 2.0000404460(1)       &   $4.04673(1)\times 10^{-5}$
\\[1pt]
$8$   & 2.00001006831(5)      &    $1.006983(5)\times 10^{-5}$
\\[1pt]
$9$   & 2.00000247685(5)       &   $2.47695(5)\times 10^{-6}$
\\[1pt]
$10$ & 2.00000060412(2)       &   $6.0412(2)\times 10^{-7}$
\\[1pt]
$11$ & 2.00000014639(1)       &   $1.4639(1)\times 10^{-7}$
\\[1pt]
$12$ & 2.000000035313(5)       &   $3.5313(5)\times 10^{-8}$
\\[1pt]
$13$ & 2.000000008489(2)       &   $8.489(2)\times 10^{-9}$
\\[1pt]
$14$ & 2.0000000020355(2)       &   $2.0355(2)\times 10^{-9}$
\\[1pt]
$15$ & 2.0000000004870(1)       &   $4.870(1)\times 10^{-10}$
\\[1pt]
$16$ & 2.00000000011634(4)       &   $1.1634(4)\times 10^{-10}$
\\[1pt]
$17$ & 2.00000000002776(2)       &   $2.776(2)\times 10^{-11}$
\\[1pt]
$18$ & 2.000000000006617(5)       &   $6.617(5)\times 10^{-12}$
\\[1pt]
$19$ & 2.000000000001575(2)       &   $1.575(2)\times 10^{-12}$
\\[1pt]
$20$ & 2.0000000000003746(8)       &   $3.746(8)\times 10^{-13}$
\\[2pt]\hline
%\noalign{\smallskip}
%\hline
%\noalign{\smallskip}
\end{tabular}
\end{center}
\label{st1}
\end{table}

%%%%%%%%%%%%%%%%%%%%%%%%%%%%%%%%%%%%%%%%%%%%
%%%%%%%%%%%%%%%%%%%%%%%%%%%%%%%%%%%%%%%%%%%%

In our work~\cite{daCosta:ddgm14}
%%, for $m=2$, $3$, and $4$, 
we found the values of $t_c$, $f(0)$, and $P(1,t_c)$ for $m=2$, $3$, and $4$ in the case of $P(1,t{=}0)=1$, that is, the initial configuration consisting of isolated nodes.

%%%%%%%%%%%%%%%%%%%%%%%%%%%%%%%%%%%%%%%%%%%%
%%%%%%%%%%%%%%%%%%%%%%%%%%%%%%%%%%%%%%%%%%%%
%%%%%%%%%%%%%%%%%%%%%%%%%%%%%%%%%%%%%%%%%%%%

\section{Above the transition}
\label{si9}

In this section we show that for $t>t_c$, where the percolation cluster is present, the evolution equations become similar to those for ordinary percolation. In the critical region near $t_c$, this enables us to perform a complete analysis of the problem using the known critical distribution as an initial condition.

Let us recall the expression of the distribution $Q(s)$ in terms of $P(s)$:
\begin{equation}
%%\nonumber
Q(s) 
%%&= \left[ 1-\sum_{u<s} P(u)\right]^m - \left[1-\sum_{u\leq s} P(u)\right]^m
%%-2P(2)
%%, \\[5pt]
%%&
= P(s)\sum_{k=0}^{m-1} \binom{m}{k+1} P(s)^k \left[1-\sum_{u \leq s}P(s)\right]^{m-1-k}
.
\label{eq:Q_P_m}
\end{equation}
%%
%where the first term on the right-hand side of first equality accounts for the contribution of the cases when all the $m$ randomly chosen clusters have sizes larger or equal to $s$, and the second term subtracts to those the cases when all the $m$ clusters have size strictly larger than $s$. 
%%where the first term on the right-hand side of first equality accounts for all the cases where all the $m$ randomly chosen clusters have sizes larger or equal to $s$, and the second term subtracts the cases where all the $m$ clusters have size strictly larger than $s$
%%This difference is equal to the fraction of cases in which at least one of the clusters has size equal to $s$ and the others have size larger than $s$, which is equivalent to the probability that a cluster selected by our rule has size $s$. 
%%
Above the percolation threshold $t_c$, where a giant component is present, the large $s$ asymptotic behavior of expression~(\ref{eq:Q_P_m}) is determined by the first term of the sum on the right-hand side (the term $k=0$)
%%equation~(\ref{e100}) 
%%(the lower power of $P(s)$), 
in which the factor $\left(1-\sum_{u\leq s} P(u)\right)^{m-1}$ can be substituted by $S^{m-1}$.  
The relation between asymptotic distributions, above $t_c$, becomes 
\begin{equation}
Q(s)\cong mS^{m-1}P(s).
\nonumber
\end{equation}

Let us introduce the generating functions of the distributions:
\begin{equation}
\rho(z) \equiv \sum_{s=1}^\infty P(s) z^s 
%\label{e30}
\label{eq:gnr_func_P_EPIAC}
\end{equation}
and 
\begin{equation}
\sigma(z) \equiv \sum_{s=1}^\infty Q(s) z^s
.
%\label{e40}
\label{eq:gnr_func_Q_EPIAC}
\end{equation}
Then for $z$ close to $1$, taking into account the normalization condition $1-\sum_s Q(s)=S^m$,  we can write the relation between generation functions~(\ref{eq:gnr_func_P_EPIAC}) and~(\ref{eq:gnr_func_Q_EPIAC}) as 
\begin{eqnarray}
&&
1-S^m-\sigma(z)=\sum_s Q(s)[1-z^s] 
\nonumber
\\[5pt]
&&
\cong \sum_s mS^{m-1}P(s)[1-z^s]=mS^{m-1}[1-S-\rho(z)]
,
%\label{eq_}
\nonumber
\end{eqnarray}
so
\begin{equation}
1-\sigma(z)=mS^{m-1}\left[1-\rho(z)-\frac{m-1}{m}S \right].
\label{eq:gnr_rlt_m}
%\nonumber
\end{equation}

Substituting the last relation into the evolution equation
\begin{equation}
\frac{\partial P(s,t)}{\partial t}
= s \sum_{u+v=s} Q(u,t)Q(v,t) - 2 sQ(s,t)
%%. 
\label{se4400}
\end{equation}
we obtain the partial differential equation for any $m$:
\begin{eqnarray}
%%\partial_t
&&
\frac{\partial\rho(z,t)
}{\partial t} 
= 
2m^2[S(t)]^{2(m-1)}
\nonumber
\\[5pt]
&&
\times\left[\rho(z,t)-1+\frac{m-1}{m}S(t)\right]
%%\partial_{\ln z}
\frac{\partial\rho(z,t)}{\partial \ln z}
.
%\label{e110}
\label{eq:gnr_pd_eq_m}
\end{eqnarray}
We use the power-law asymptotics of the distribution $P(s, t_c) \cong f(0)s^{1-\tau}$ as the initial
condition for Eq.~(\ref{eq:gnr_pd_eq_m}). This corresponds to the following singularity of the generating function at $z=1$:
\begin{equation}
1-\rho(z,t_c) = \text{analytic terms}-f(0)\Gamma(2-\tau)(1-z)^{\tau-2}
.
\label{se4600}
\end{equation}
%%
%%To check the continuity of the transition, one again 
We substitute $S(t)=B(t-t_c)^\beta$ into Eq.~(\ref{eq:gnr_pd_eq_m}), and rewrite it in terms of the transformed variables $\epsilon\equiv(t-t_c)^{(m-1)2\beta+1}$ and $x\equiv\ln z$:
\begin{equation}
\frac{\partial\rho}{\partial\epsilon} = \frac{2m^2B^{2(m-1)}}{1+(m-1)2\beta}\Bigg(\rho-1+\frac{m-1}{m}B\epsilon^{\beta/[1+(m-1)2\beta]}\Bigg)\frac{\partial\rho}{\partial x}
. 
%\label{e130}
\label{eq:gnr_pd_eq_2_m}
\end{equation}
To solve this equation, we use the hodograph transformation approach. We pass from $\rho = \rho(x,\epsilon)$ to $x = x(\rho,\epsilon)$, which leads to a simple linear partial differential equation for $x(\rho,\epsilon)$ and
enables us to find the general solution
\begin{eqnarray} 
&&
\ln z = \frac{2m^2B^{2(m-1)}}{1+(m-1)2\beta}
\nonumber
\\[5pt]
&&
\times\Bigg[1-\rho-\frac{m-1}{m}B\frac{(t-t_c)^\beta}{1{+}\beta/[1{+}(m-1)2\beta]}\Bigg](t-t_c)^{1+(m-1)2\beta} 
\nonumber
\\[5pt]
&&
+ F(\rho)
, 
\label{se4800}
%\label{e150}
\end{eqnarray}
where the function $F(\rho)$ is obtained from the initial condition (\ref{se4600}), which gives the solution:
%%The solution of this partial differential equation is obtained with the application of the same technique used above. We take the critical distribution as initial condition, thus getting $\rho(t,z\to 1)$ implicitly, for the critical region above $t_c$ (when there is a giant component):
%%
\begin{eqnarray} 
&&
\ln z = \frac{2m^2B^{2(m-1)}}{1+(m-1)2\beta}
\nonumber
\\[5pt]
&&
\times\Bigg[1-\rho-\frac{m-1}{m}B\frac{(t-t_c)^\beta}{1{+}\beta/[1{+}(m-1)2\beta]}\Bigg](t-t_c)^{1+(m-1)2\beta} 
\nonumber
\\[5pt]
&&
-[f(0)]^{-1/(\tau-2)}|\Gamma(2-\tau)|^{-1/(\tau-2)}[1-\rho]^{1/(\tau-2)}
. 
\label{eq:gnr_sol_m}
%\label{e150}
\end{eqnarray}
Setting $z=1$ and taking into account the relation $1-\rho(t,1)=S(t)=B(t-t_c)^\beta$ and comparing resulting
powers and coefficients in Eq.~(\ref{eq:gnr_sol_m}), we obtain relations between critical exponents 
%%Accounting for the identity $1-\rho(t,1)=S(t)$, inspection of equation~(\ref{eq:gnr_sol_m}) at $z=1$ (where it is exact) shows that the assumed continuous variation of giant component size ($S(t)=B(t-t_c)^\beta$) is indeed a solution of the evolution equation~(\ref{e200}) under selection rule~(\ref{eq:Q_P_m}). This equation provides us with the relation between critical exponents $\tau$ and $\beta$:
%%
\begin{equation}
\tau=2+\frac{\beta}{1+(2m-1)\beta},
\label{eq:tau_beta_m}
\end{equation}
 and between critical amplitudes $B$ and $f(0)$:
% \begin{equation}
%B = \Bigg[2m\frac{[1-(2m-1)(\tau-2)][1+(m-1)(\tau-2)]}{3-\tau}\Bigg]^{(\tau-2)/[1-(2m-1)(\tau-2)]} \left(f(0) |\Gamma(2-\tau)|\right)^{1/[1-(2m-1)(\tau-2)]}
%\label{eq:B_f(0)_m}
%\end{equation}
%%
 \begin{eqnarray}
&&
B = \left[  f(0) |\Gamma(2-\tau)|\right]^{1/[1-(2m-1)(\tau-2)]} 
\nonumber
\\[5pt]
&&
\!\!\!\!\!\!\!\!\!\!\!\!\!\!\!\!\!\!
\times\!\!
%%\left[\left(2m\frac{[1-(2m-1)(\tau-2)][1+(m-1)(\tau-2)]}{3-\tau}\right)^{\tau-2}\right]^{1/[1-(2m-1)(\tau-2)]}
\left[\!2m\frac{[1{-}(2m{-}1)(\tau{-}2)][1{+}(m{-}1)(\tau{-}2)]}{3-\tau}\!\right]^{\!\textstyle\frac{\tau-2}{1-(2m-1)(\tau-2)}}
%%{(\tau-2)/[1-(2m-1)(\tau-2)]}
\!\!\!\!\!\!\!\!\!\!\!\!\!\!\!\!\!
,
\label{eq:B_f(0)_m}
\end{eqnarray}
for an arbitrary $m$. 

One can easily show that 
%%above the transition, 
Eqs.~(\ref{se2800}) and (\ref{se2900}) for the scaling functions $\tilde{f}(x)$ and $\tilde{g}(x)$ derived for the normal phase are also valid for the percolation phase ($t>t_c$) after the following modification: one must invert signs of each term of Eq.~(\ref{se2800}) which contains $\tilde{f}(x)$ or its derivatives. The same applies to the equation for $\tilde{f''}(x)$ in system (\ref{se3300}). 
Then, similarly to the normal phase, we find that 
%%
%%It is clear that the effectiveness of the selection rule's bias is determined by the model parameter $m$, a 
%%A larger $m$ leads to a more efficient selection of small clusters. 
%%Thus, the previous analysis proves that the explosive percolation transition is always continuous, even for an arbitrarily strong bias, as long as the number of comparisons at each step remains finite (i.e. $m$ is finite). 
%%When $m$ approaches infinity there is indeed a discontinuity at the limiting point $t_c=1$ at which $S$ jumps from $0$ to $1$. 
%%, since the dynamics simply merges the two smallest clusters present at each step.
%%Insets on Fig.~\ref{f4} highlight the behavior of scaling functions near the origin. We find the singularities on $f(x)$ and $g(x)$ at $x=0$. The top inset shows the derivatives of the curves plotted in the main panel for the same range of $x$, and the bottom one shows a zoom of the near $x=0$ region. The 
the derivatives of $f(x)$ and $g(x)$ diverge approaching $x=0$ also from above. 
%%, from both sides. 
This singular behavior is 
%%correctly predicted 
described by series~(\ref{se3000}), that were written for the disordered phase, but, in fact, hold on both sides of the transition. It is clear that $f(0)$ and $g(0)$ should be equal on both phases. Moreover, the series coefficients of $f(x)$ and $g(x)$ above $t_c$, can be found similarly to 
%%in a similar way as for 
$t<t_c$.  
%%In the percolation phase, the coefficients $a_1,b_1,a_2,b_2\dots$ have the same moduli as the corresponding coefficients in the normal phase but opposite signs. 
%%That is, in the percolation phase, $-a_1,-b_1,-a_2,-b_2\dots$ are given by the right-hand sides of expressions~(\ref{se3700}).
%%In the percolation phase, the coefficients $a_k$ and $b_k$ have the same moduli as the corresponding coefficients in the normal phase, and the only difference is on their signs. Namely,  in the percolation phase the coefficients are given by Eqs.~(\ref{se3700}) after the following transformation $a_k \to (-1)^k a_k$ and $b_k \to (-1)^k b_k$.
In the percolation phase, the coefficients $a_k$ and $b_k$ are given by Eq.~(\ref{se3700}), which was derived for the normal phase, after the following transformation $a_k \to (-1)^k a_k$ and $b_k \to (-1)^k b_k$.

%%%%%%%%%%%%%%%%%%%%%%%%%%%%%%%%%%%%%%%%%%%%
%%%%%%%%%%%%%%%%%%%%%%%%%%%%%%%%%%%%%%%%%%%%
%%%%%%%%%%%%%%%%%%%%%%%%%%%%%%%%%%%%%%%%%%%%

\section{Power-law initial conditions} 
\label{si10}

Here we show that if the initial size distribution of clusters decays sufficiently slowly, the transition takes place at the initial moment. 

Let us assume that the initial distribution is power-law, $P(s,t=0) \sim s^{1-\tau_0}$, where the exponent $\tau_0$ defines the initial condition. This distribution results in divergent susceptibility if, according to Eq.~(\ref{e1800})  and Eq.~(\ref{os6}), 
\begin{equation}
\int_{\text{const}}^{\infty} \!\!\!\! ds\, [s^{(2m-1)-m\tau_0}]^2/s^{-\tau_0} 
%%\sim \int_{const}^{\infty} \!\!\!\! ds\, s^{2(2m-1)-(2m-1)\tau_0} 
= \infty
, 
\label{e3x00}
\end{equation}
that is if  
\begin{equation}
\tau_0 \leq 2+1/(2m-1)  
. 
\label{e3y00}
\end{equation}
The divergent susceptibility indicates the presence of the continuous transition exactly at the point of divergence.  
So, if this condition is satisfied, then the transition occurs at the initial instant, i.e. $t_c=0$. 
We will describe this case in detail elsewhere. 
%%In this case, we have $\tau=\tau_0$ and all other exponents can be found from equations~(\ref{e1400})--(\ref{e1700}) by substituting $\tau=\tau_0$. For example, 
%%
%%\begin{equation}
%%\beta = \frac{\tau_0 - 2}{2 + 1/(2m-1) - \tau_0}
%%\beta = \frac{\tau_0 - 2}{1 - (2m-1)(\tau_0-2)}
%%. 
%%\label{e3z00}
%%\end{equation}
%%
On the other hand, if $\tau_0>2+1/(2m-1)$, then we 
%%have the transition point $t_c>0$, and 
arrive at the situation described 
%%have the exponents which we have obtained 
in the previous sections, namely, $t_c>0$ ($t_c$ depends on $\tau_0$), and the critical exponent values (independent of $\tau_0$) presented in Table~\ref{st1}.

%%%%%%%%%%%%%%%%%%%%%%%%%%%%%%%%%%%%%%%%%%%%
%%%%%%%%%%%%%%%%%%%%%%%%%%%%%%%%%%%%%%%%%%%%
%%%%%%%%%%%%%%%%%%%%%%%%%%%%%%%%%%%%%%%%%%%%

\section{Simple estimates for the percolation threshold} 
\label{si11}

Let us estimate $t_c$ in the case of $m=2$  
assuming that the process starts from isolated nodes, i.e., \mbox{$P(1,t=0)=1$}. 
%%In this case, the distribution $P(s,t_c)$ is close to its power-law asymptotics $P(s\gg 1,t_c) \cong f(0)s^{1-\tau}$ even in the range of small $s$. 
%%Let the process start from isolated nodes 
%%As $m$ tends to infinity, $\tau$ and $t_c$ approach $2$ and, respectively, $1$. Our theory provides the critical exponents and scaling functions. To find $t_c$, 
%%which is determined by 
%%for a given initial cluster size distribution, in principle, one has to solve numerically the set of evolution equations (\ref{e200}). This was made in Ref.~\cite{daCosta:ddgm10} in the particular case of $m=2$. 
%%For a given initial cluster size distribution, one can estimate 
%%However, if $m$ is not very large, we can estimate $t_c$ without solving the master equations numerically. 
%%even without solving the master equations numerically, we can 
%%%%easily 
%%estimate $t_c$ if $m$ is not very large. 
%%%%much greater than $1$. 
%%Furthermore, for a given initial distribution, one can relate critical exponent values and $t_c$. In particular, if $m=2$, at the initial instant all the nodes be isolated, we can make a simple estimate. 
The numerical solution of evolution equations for $P(s,t)$ showed that 
for sufficiently small $m$, including $m=2$, the asymptotic power-law at the critical point, $P(s,t_c) \cong f(0)s^{1-\tau}$, is still approximately valid even at small $s$, and, moreover, $f(0)$ deviates from $P(s=1,t_c)$ only by a small number of the order of $\tau-2$ if all nodes initially were isolated. In this special case, we can approximate $P(s,t_c)$ in the sum rule $\sum_{s=1}^\infty P(s,t_c) = 1$ by $P(s=1,t_c)s^{1-\tau}$ at any $s\geq 1$, which gives 
\begin{equation}
P(1,t_c)\zeta(\tau-1) \approx 1, 
\label{e3400}
\end{equation}
where $\zeta(x) \equiv \sum_{s{=}1}^\infty s^{-x}$ is the Riemann zeta function. We 
%%can 
find $P(1,t)$ explicitly in the full range of $t$ by solving the master Eq.~(\ref{e200}) with the initial condition $P(1,0)=1$. Let, e.g., $m=2$. Then the result is
\begin{equation}
P(1,t) = \frac{2}{1 + e^{4t}} ,
\label{e3500}
\end{equation}
so we have 
\begin{equation}
\frac{2}{1 + e^{4t_c}}\zeta(\tau-1) \approx 1,
\label{e3600}
\end{equation}
and finally 
\begin{equation}
t_c \approx \frac{1}{4}\ln[2\zeta(\tau-1)-1].
\label{e3700}
\end{equation}
Substituting $\tau=2.04763044$, which we obtained above for $m=2$ into this formula, we finally find an estimate for $t_c$, namely $t_c \approx 0.935$. This estimate is close to a precise value $t_c=0.92320750930(2)$ obtained in our work~\cite{daCosta:ddgm14}.

%%\newpage

%%%%%%%%%%%%%%%%%%%%%%%%%%%%%%%%%%%%%%%%%%%%%%%%%%%
%%%%%%%%%%%%%%%%%%%%%%%%%%%%%%%%%%%%%%%%%%%%%%%%%%%
%%%%%%%%%%%%%%%%%%%%%%%%%%%%%%%%%%%%%%%%%%%%%%%%%%%
%%%%%%%%%%%%%%%%%%%%    B I B L I O G R A P H Y    %%%%%%%%%%%%%%%%%
%%%%%%%%%%%%%%%%%%%%%%%%%%%%%%%%%%%%%%%%%%%%%%%%%%%
%%%%%%%%%%%%%%%%%%%%%%%%%%%%%%%%%%%%%%%%%%%%%%%%%%%
%%%%%%%%%%%%%%%%%%%%%%%%%%%%%%%%%%%%%%%%%%%%%%%%%%%

%%\begin{thebibliography}{10}

\end{document}